\documentclass[aps,prd,onecolumn,nofootinbib,superscriptaddress,12pt]{revtex4}
\pdfoutput=1
\usepackage{float}
\usepackage{graphicx}
\usepackage{amsmath}
\usepackage{empheq}
\usepackage{amsfonts}
\usepackage{amssymb,ulem}
\usepackage{color}%
\usepackage{dcolumn}
\usepackage{amsmath,amssymb,mathrsfs}
\usepackage{slashed}
\usepackage{subfigure}
\usepackage{multirow}
\usepackage{csquotes}

\usepackage{MnSymbol,wasysym}
\usepackage{braket,diagbox}
\usepackage{eurosym}
\usepackage[usenames,dvipsnames,svgnames]{xcolor}

\newcommand{\RNum}[1]{\uppercase\expandafter{\romannumeral #1\relax}}
\usepackage[colorlinks=true,linkcolor=blue,urlcolor=blue,filecolor=black,citecolor=red,
pdfstartview=FitV,pdftitle={},pdfsubject={},pdfkeywords={},pdfpagemode=None,bookmarksopen=true]{hyperref}

\usepackage{pifont}
\usepackage{booktabs}

\begin{document}
\baselineskip=0.6 cm

\title{Gauge Symmetries, Exact Symmetries  and
Conserved Charges in Minimal Massive Gravity}

\author{Kang Liu}
\email{liukanyang@sina.com}
\affiliation{\baselineskip=0.4cm Center for Gravitation and Cosmology, College of Physical Science and Technology, Yangzhou University, Yangzhou, 225009, China}

\author{Xiao-Mei Kuang}
\email{xmeikuang@yzu.edu.cn}
\affiliation{\baselineskip=0.4cm Center for Gravitation and Cosmology, College of Physical Science and Technology, Yangzhou University, Yangzhou, 225009, China}

\begin{abstract}
\baselineskip=0.6 cm
In this paper, we investigate a three-dimensional gravitational model known as Minimal Massive Gravity (MMG), which includes an auxiliary field, using the covariant phase space method. Our analysis reveals the presence of three gauge symmetries whose algebras close via field recombination and parameter classification within this framework. Upon incorporating these additional symmetries within a specific limit of parameters, we find that the Kosmann derivative should be replaced by a novel transformation compatible with Wald's approach, which establishes a new mechanism for generating exact symmetries and constructing their corresponding conserved charge  in theories with auxiliary fields, extending beyond standard methods. However, this transformation does not yield closed algebras on the space of fundamental fields. We find that this corresponds to a Lorentz vector that characterizes the approximate completeness of translation symmetry. As a result, we obtain a gauge invariant charge at a certain limit of parameters, which emerges as a nontrivial combination of the diffeomorphism charge and integrable gauge charges.

\end{abstract}

\maketitle
\tableofcontents

\section{Introduction}
Developing a quantum gravity theory remains a persistent challenge that combines technical complexity with unresolved conceptual gaps. Notably, standard Einstein-Hilbert gravity in three-dimensional spacetime is topological in the sense
that it does not possess local degrees of freedom, which allows exact solutions through Chern-Simons formulations with gauge group $\rm{SL(2) \times SL(2)}$ \cite{Achucarro:1986uwr,Witten:1988hc}. This topological reduction serves as both computational laboratory and conceptual prototype for quantum gravity research.

The incorporation of a gravitational Chern-Simons term into the action principle generates propagating gravitational excitations, thereby establishing a distinct theoretical framework known as Topologically Massive Gravity ($\rm{TMG}$) \cite{Deser:1982vy}. This theory exhibits unexpected stability at the critical point $\mu\ell=1$  \cite{Li:2008dq}, where $\ell$ denotes the $\rm{AdS}$ radius, initially interpreted as the emergence of a chiral structure. Further studies reveal that even at this specific chiral point, certain left-moving excitations persist \cite{Li:2008dq, Witten:1988hc, Maloney:2007ud, Carlip:2008jk, Grumiller:2008qz, Skenderis:2009nt, Blagojevic:2008bn, Giribet:2008bw}. This theory adopts a relaxed version of Brown-Henneaux boundary conditions but follows Grumiller-Johansson fall-off conditions instead \cite{Grumiller:2008qz, Grumiller:2008es, Henneaux:2009pw}.

A broader framework of three-dimensional gravity is $\rm{MMG}$, first proposed in \cite{Bergshoeff:2014pca}. By introducing a curvature-squared term modification within a specific parameter range, this theory constructs a minimal gravitational model,  which retains TMG's single spin-2 excitation in the bulk but resolves its unitarity paradox and simultaneously remains the energies of bulk graviton to be positive \cite{Arvanitakis:2014xna, Alkac:2017vgg, Bergshoeff:2009aq}. This resolution is achieved through parameter regimes with branches of positive central charges ${c_ \pm } = \frac{{3\ell }}{{2{G_3}}}\left( {\sigma  \pm \frac{1}{{\mu \ell }} + \alpha C} \right)$, where $G_3$ is three-dimensional Newton constant and $\sigma, \alpha, C$ are other parameters in the model. In general, central charges emerge from the central term in the Virasoro algebra, which arises from the gravitational anomaly of dual $\rm{2d}$ conformal field theory. The Hamiltonian in $\rm{MMG}$ analysis is set in the separated time and space components of the Lorentz-vector valued one-form with Poisson brackets algebra \cite{Bergshoeff:2014pca,Hohm:2012vh,Bergshoeff:2014bia,MahdavianYekta:2015tmp}, which helps to calculate the central charge ${c_\pm }$. The No-tachyon and No-ghost conditions require the parameters to be in ranges of $2C<1$ and $ \pm \sigma (1 + \alpha \sigma ) < 0$. The linearized equations in $\rm{MMG}$ can be expressed by quadratic Casimir operators in $\rm{SL(2) \times SL(2)}$ algebra to get left and right moving massless gravitons, and they also allow the logarithmic modes at a new critical point \cite{Alishahiha:2014dma}. In recent years, the influence of parameter ranges on the gravitational theory has also been extensively studied \cite{Hajian:2015xlp, Tavlayan:2024zbl, Deger:2023eah, Xiao:2023lap,Hajian:2023bhq}. Thus, MMG can be utilized as an ideal model to determine the existence of potential new symmetry, identify its corresponding conserved charges and exact symmetries, and establish its relationship with the theory's parameters. A similar issue has been considered in the Mielke–Baekler (MB) model \cite{Geiller:2020edh,Banerjee:2012jn,Banerjee:2009vf,Banerjee:2011cu}, but  the corresponding  study in MMG is still  lacking.

Cartan formalism employs a triad one-form $e$ and a connection one-form $\omega$ as gauge fields, reinterpreting gravity as a gauge theory of the Lorentz group \cite{Yilmaz:2007ij,Keurentjes:2002xc,DePaoli:2018erh,Jubb:2016qzt,Wieland:2017zkf,Oliveri:2019gvm, Frodden:2019ylc}. These gauge transformations decompose into two distinct classes: Lorentz-type transformations of frame fields and translational symmetries. They are the same as the infinitesimal version of ordinary $\rm{ISO(2,1)}$ gauge transformation of the Chern-Simons one-form field $A$ \cite{Kiran:2014dfa}. However, such formulations manifestly break general covariance under $\rm{ISO}(3,1)$ (Poincar$\acute{e}$), $\rm{SO}(3,2)$ ($\rm{AdS}$) or $\rm{SO}(4,1)$ ($\rm{dS}$), as evidenced by the absence of four-dimensional diffeomorphism-invariant structures in first-order formulations \cite{Zanelli:2005sa}. The approach of Wald \cite{Wald:1993nt,Frodden:2019ylc, Jacobson:2015uqa} and Hamiltonian analysis \cite{Cacciatori:2005wz,Giacomini:2006dr,Banerjee:2009vf,Banerjee:2011cu} can also be generalized into the Cartan formalism. In metric formalism, equations of motions of $\rm{MMG}$ cannot be derived from an action containing only the metric field, and similar to  TMG, the solutions of MMG have Segre-Petrov types $\text N$ and $\text D$ after redefinitions of parameters \cite{Charyyev:2017uuu}. Additionally,  it was found that in holographic renormalization, MMG yields an identical boundary theory in the limits of $(\alpha=0,\mu\to\infty)$ and $(\alpha\ne0,\mu\to\infty)$,  suggesting that the terms proportional to $\alpha$ do not affect the theory's asymptotics \cite{Alishahiha:2015whv}.

In stationary background, bifurcation surface  ${\cal B}_H$ is a part of the co-dimension two surface $\cal S$ and is characterized by the vanishing Killing vector on it, accompanied by a non-zero and constant surface gravity $\kappa$. Moreover, on $\cal B$, the covariant derivative of a Killing vector $\xi_K$ satisfies ${\nabla ^\mu }{{\xi_K} ^\nu } = \kappa {n^{\mu \nu }}$, where $n^{\mu\nu}$ is the binormal to $\cal B$       \cite{Wald:1999wa,Poisson:2009pwt}. For Wald's method \cite{Wald:1993nt}, the black hole entropy arises from the replacement of $\kappa n^{\mu\nu}$ by ${\nabla ^\mu }{{\xi_K} ^\nu }$ in the existence of exact symmetries. Moreover, the Lorentz-Lie derivative, or Kosmann derivative \cite{Jacobson:2015uqa}, leads to a general formula for black hole entropy \cite{,Setare:2015nla,Adami:2015onz,Setare:2015gss,Setare:2017wuj,Adami:2017phg}. More new developments in this direction have been addressed in dynamical background or other higher derivative theories \cite{Hollands:2024vbe, Kong:2024sqc, Jiang:2018sqj,Guo:2024oey}.

The main aim of this paper is to investigate the gauge symmetries and their associated conserved charges in MMG using the covariant phase space method.  MMG is formulated as a first-order theory where the triad, spin-connection, and a new degree of freedom serve as the fundamental fields. For this class of fields, it is well established that Lorentz and translational symmetries correspond to the fundamental internal symmetries, but their related  algebras were found not to be closed in the solution space \cite{MahdavianYekta:2015tmp, Boulanger:2023tvt}, which implies a lack of algebraic completeness. This raises a natural set of questions: can the new degree of freedom introduce new symmetry? If so, what are its corresponding conserved charges and exact symmetries? Moreover, what is the relationship between the existence of such a symmetry and the selection of the theory's parameters?  We believe that MMG can serve as a proper theory model  to investigate the relationship between a theory's parameters and its symmetries.

This paper is organized as follows. In Section \ref{basis} , we first introduce the foundational elements of $\rm{MMG}$, then we  investigate the existence conditions for gauge symmetries in this theory and derive the transformations whose algebras close. Our results  include new transformations and algebras associated with the auxiliary field $h$. In Section \ref{section 3}, we incorporate three types of gauge charges and diffeomorphism charge with a fixed integrable charge. In Section \ref{section 4}, based on the considerations regarding the Lie derivatives of certain field components and constraints on the parameters of $\rm{MMG}$, we construct a new transformation satisfying approximately closed algebras on an extended field space within a specific limit of parameters. This transformation, when performing on the new field, is also shown to generate exact symmetries, which depend on the computation of specific Lorentz vectors. This property enables gauge symmetries  to generate exact symmetries and a generalization of conserved charges. We also verify our proposal in a specific example. The last section is our conclusions and discussions.
In addition, in Appendix \ref{appendix one} we give the notation and convention used in this study, and Appendixes \ref{appendix 3} and \ref{appendix c} contain detailed proofs of some calculations.

\section{Gauge symmetries in MMG}
\label{basis}
\subsection{Basis of $\rm{MMG}$}
The theory of MMG features three flavors of fields: one-form triad $e$, dual one-form spin connection $\omega$ and an auxiliary one-form field $h$.  This theory postulates the following equations of motion \cite{Bergshoeff:2014pca}
\begin{align}\label{eommetric}
	\frac{1}{\mu} C_{\mu \nu}+\bar{\sigma} G_{\mu \nu}+\bar{\Lambda}_0 g_{\mu \nu}=-\frac{\gamma}{\mu^2} J_{\mu \nu},
\end{align}
where $\bar{\sigma}, \bar{\Lambda}_0, \gamma, \mu$ are constants,  $g_{\mu\nu}$ is the metric tensor and $G_{\mu\nu}$ is the related  Einstein tensor. $C_{\mu\nu}$ is Cotton tensor defined by Schouten tensor $S_{\mu\nu}=R_{\mu\nu}-\frac{1}{4}R g_{\mu\nu}$  as
\begin{align}
	{C_{\mu \nu }} = \frac{1}{{\sqrt { \left| g \right|} }}{\tilde\varepsilon _\mu }^{\alpha \beta }{\nabla _\alpha }{S_{\beta \nu }}
\end{align}
with $\tilde\varepsilon_{012}=1$.
The tensor $J_{\mu\nu}$  is defined as
\begin{align}
	J_{\mu\nu}=R_\mu{ }^\rho R_{\rho \nu}-\frac{3}{4} R R_{\mu \nu}-\frac{1}{2} g_{\mu \nu}\left(R^{\rho \sigma} R_{\rho \sigma}-\frac{5}{8} R^2\right),
\end{align}	
which is covariant conserved when the equation of motion \eqref{eommetric} is satisfied. It is noted that this gravity theory has no metric formalism of Lagrangian, but it has an equivalent one on field space $(\omega, e, h)$
\begin{align}\label{main L}
	&L =  - \sigma e \wedge R(\omega ) + \frac{{{\Lambda _0}}}{6}e \wedge [e \wedge e] + h \wedge T(e) + \frac{1}{{2\mu }}\left( {\omega \wedge \text{d}\omega  + \frac{1}{3}\omega  \wedge [\omega  \wedge \omega] } \right)\notag\\
	& + \frac{\alpha }{2}e \wedge [h \wedge h],
\end{align}
where $[\cdot \wedge \cdot]$ represents the Lie algebra commutator, and $\sigma$, $\alpha$ and $\Lambda_0$ are additional parameters.  $T(e)$ and $R(\omega)$ are locally covariant torsion and curvature two-forms, which are respectively defined as
\begin{align}\label{T,R}
	T(e )=\text{d}_{\omega}e = \text{d}e + \left[ {\omega  \wedge e} \right], \qquad R(\omega ) = \text{d}\omega  + \frac{1}{2}\left[ {\omega  \wedge \omega } \right].
\end{align}	
The fields $e$, $\omega$ and $h$ combine the even parity objects $T(e)$ and $R(\omega)$ together to give the three-forms Lagrangian $L$. Readers can refer to Appendix \ref{appendix one} for clearly tracing the notations and operations  in the Lagrangian. Specially, the field $h$ corresponds to a new degree of freedom in the theory, also known as an auxiliary field \cite{Merbis:2014vja, Bergshoeff:2009zz, Hohm:2012vh, Bergshoeff:2013xma}, which essentially is a new field introduced to fix the derivative of $\omega$ \cite{Hohm:2012vh, Blagojevic:2010ir}.

The variation of $L$ gives three equations of motion ${E_h},{E_\omega}$ and ${E_e}$  and the symplectic potential $\theta$ current,
\begin{align}\label{L}
	\delta L = {E_h} \wedge \delta h + {E_\omega } \wedge \delta \omega  + {E_e} \wedge \delta e + \text{d}\theta ,
\end{align}	
where
\begin{subequations}\label{EOM}
	\begin{align}
		& E_h=\frac{{\partial L}}{{\partial h}} =T(\omega)+\alpha \left[ {e \wedge h} \right]\label{eom21}, \\
		& E_\omega=\mu({\text{d} }\frac{{\partial L}}{{\partial {\text{d} }\omega }} - \frac{{\partial L}}{{\partial \omega }})=R(\omega)+\mu \left[ {e \wedge h} \right]-\sigma \mu T(\omega)\label{eom22}, \\
		& E_e={\text{d} }\frac{{\partial L}}{{\partial \text{d} e}} - \frac{{\partial L}}{{\partial e}} =-\sigma R(\omega)+\frac{\Lambda_0}{2} \left[ {e \wedge e} \right]+{\text{d}}_\omega h+\frac{\alpha}{2} \left[ {h \wedge h} \right],\label{eom23}
	\end{align}	\label{eomtwo}
\end{subequations}
and
\begin{align}
	\theta  = \delta \omega  \wedge ( - \sigma e + \frac{1}{{2\mu }}\omega ) + \delta e \wedge h\label{symplectic potential}.
\end{align}
 By combining the equations in  \eqref{EOM}, one finds that  $E_e$ coincides with  \eqref{eommetric}  when $1+\sigma\alpha\ne0$, and the parameters satisfy
\begin{align}\label{parameters relations}
\begin{aligned}
	 \bar \sigma  = \sigma  + \alpha \left[ {1 + \frac{{\alpha {\Lambda _0}/{\mu ^2}}}{{2{{(1 + \sigma \alpha )}^2}}}} \right],~ \gamma  =  - \frac{\alpha }{{{{(1 + \sigma \alpha )}^2}}},~ {{\bar \Lambda }_0} = {\Lambda _0}\left[ {1 + \sigma \alpha  - \frac{{{\alpha ^3}{\Lambda _0}/{\mu ^2}}}{{4{{(1 + \sigma \alpha )}^2}}}} \right].
    \end{aligned}
\end{align}
Moreover, seeking  the solutions to \eqref{EOM}  for  $1+\sigma\alpha=0$ is also important and well motivated. On one hand,  this choice of parameter is closely related with the closure of algebras, which will be studied  in Section \ref{section two}.  On the other hand, by introducing a torsion free spin-connection $\Omega$,
\begin{align}\label{second connection}
	\Omega=\omega+\alpha h\quad \quad \mathrm{or} \quad \quad {\Omega_\mu}^I={\omega_\mu}^I+\alpha {h_\mu}^I,
\end{align}
one can have two other types of two-forms: $T(\Omega)=\text{d}_\Omega e=0$ and $R(\Omega )=d\Omega+\frac{1}{2}\left[ {\Omega \wedge \Omega} \right]$.  Subsequently,  \eqref{eom23} can be reduced into
\begin{align} \label{equation one of h}
	{\text{d}_\Omega }h - \frac{\alpha }{2}\left[ {h \wedge h} \right] + \sigma \mu (1 + \sigma \alpha )\left[ {e \wedge h} \right] + \frac{{{\Lambda _0}}}{2}\left[ {e \wedge e} \right] = 0,
\end{align}
which suggests that  $h$ has an on-shell formula
\begin{align}\label{first solution of h}
	{h_{\mu \nu }} = {h_\mu }^I{e_{\nu I}} =  - \frac{1}{{\mu {{(1 + \alpha \sigma )}^2}}}\left( {{S_{\mu \nu }} + \frac{{\alpha {\Lambda _0}}}{2}{g_{\mu \nu }}} \right).
\end{align}
It is noted that the solution  \eqref{first solution of h} is acceptable only when $1+\sigma\alpha\ne0$. When $1+\sigma\alpha=0$, \eqref{eom23} can be reduced into
\begin{align}\label{eoms}
	{S_{\mu \nu }} =- \frac{{\alpha {\Lambda _0}}}{2}{g_{\mu \nu }},
\end{align}
and the solution of ${h_\mu}^I$ from \eqref{equation one of h} is
\begin{align}\label{second solution of h}
	h =C_1 e,
\end{align}
where ${C_1} =  \pm \sqrt {\left| {\frac{{{\Lambda _0}}}{\alpha }} \right|} $. It is noted that  ${h_\mu}^I$ is similar to the solution found in \cite{Tekin:2014jna, Alishahiha:2015whv} in asymptotic $\rm{AdS}$ region  or the maximally symmetric solution \cite{MahdavianYekta:2015tmp} for $1+\sigma\alpha\ne0$. The point $1+\sigma\alpha=0$ is actually where the Chern-Simons-like theory loses degree of freedom \cite{Witten:1988hc}. Even so, we will verify in Section \ref{section 4} that the exact symmetry in our context is still supported in the limit $1+\sigma\alpha\to0$. The new field $h$ will also introduce novel properties of $\rm{MMG}$, including the gauge symmetries, closed algebras and charges, as will be shown in subsequent analysis.

\subsection{Gauge symmetries}
\label{section two}
In this subsection, we shall investigate the gauge symmetries and their closed algebras for the $\rm{MMG}$. The main operators used in the current paper are displayed in Table \ref{operators}, including a novel contraction $I_\chi^\text{m}$ and a new translation $\delta_\chi^\text{m}$ associated with the auxiliary field $h$.  Before proceeding, we shall give a brief introduction on the gauge symmetries.  For a $d$-form Lagrangian $L(\Phi,\partial \Phi,\dots)$ and a transformation $\delta_f \Phi=f$,  if this transformation satisfies
\begin{align}\label{definition of tansformation}
	\delta_f L=\text{d} B_f,
\end{align}
where $B_f$ is a $(d-1)$-form, then $f$ is a gauge  symmetry of the Lagrangian theory. Though two gauge symmetries $f_1$ and $f_2$ differ by a  transformation $f(\alpha)$ , i.e.,  $f_2=f_1+f(\alpha)$, where $\alpha$ is an arbitrary function, such that  ${\delta _{{f_2}}}L - {\delta _{{f_1}}}L \approx \text{d}{B_f}$ due to the on-shell condition, they are on-shell equivalent.
\begin{table}[htbp]
	\centering
	\caption{Main operators used in this article}
	\label{operators}
	\begin{tabular}{lccc}
		\toprule
		\midrule
        \hline\hline
		$i_\xi$     &  Contraction along vector $\xi$&	$\text{d}$ & Exterior differential derivative  \\ \hline
		$\delta$& Variation of phase space&${\cal L}_\xi$& Lie derivative\\
        \hline
       $I_\tau^{\text{j}}$ & Contraction of the Lorentz transformation&$\delta_\tau^{\text{j}}$ & Lorentz transformation  \\
       \hline
		$I_\rho^{\text{t}}$ & Contraction of the translation&$\delta_\rho^{\text{t}}$ & Translation\\
        \hline
		$I_\chi^{\text{m}}$ & Contraction along auxiliary direction&$\delta_\chi^{\text{m}}$ & New translation introduced by auxiliary field\\
        \hline
         ${\cal K}_\xi$& Kosmann derivative relating with $\delta_\varphi^{\text{j}}$ &$\delta_{(\xi,\beta_\xi,\bar\beta_\xi)}$& New transformation generating exact symmetries\\
         \hline\hline
		\bottomrule
	\end{tabular}
\end{table}

In order to derive the gauge symmetries in $\rm{MMG}$, we construct transformations generated by spacetime diffeomorphisms performing on an arbitrary form $X$ via a vector $\xi$, following the Cartan formula ${\cal L}_\xi X=\text{d} i_\xi X+i_\xi \text{d}X$. By considering the equations of motion \eqref{eomtwo},  ${\cal L}_\xi$ performs on $e$, $\omega$, and $h$ yielding
	\begin{align}\label{eq:Le-e}
		& {\cal L}_\xi e=\text{d}i_\xi e+i_\xi \text{d}e \notag \\
		& \approx \text{d}\rho+i_\xi(-\alpha[e\wedge h]-[\omega\wedge e]) \notag \\
		& =\text{d}_\omega\rho+\alpha\left[h\wedge\rho\right]+\alpha\left[e\wedge\chi\right]+\left[e\wedge\tau\right],
\end{align}
            \begin{align} \label{eq:Le-omega}
		& {\cal L}_\xi\omega=\text{d}i_\xi\omega+i_\xi\text{d}\omega   \notag\\
        &\approx \text{d}\tau-[\tau\wedge\omega]-\mu[\rho\wedge h]+\mu[e\wedge\chi]-\sigma\mu\alpha([\rho\wedge h]-[e\wedge\chi])  \notag\\
		& =\text{d}_{\omega}\tau+(1+\sigma\alpha)\mu\left([h\wedge\rho]+[e\wedge\chi]\right),
	\end{align}

and
\begin{align}\label{eq:Le-h}
	\begin{aligned}
		& {\cal L}_{\xi}h=\text{d}i_{\xi}h+i_{\xi}\text{d}h \\
		& \approx \text{d}\chi-\sigma\mu(1+\sigma\alpha)\left(\left[\rho\wedge h\right]-\left[e\wedge\chi\right]\right)-\Lambda_0\left[\rho\wedge e\right] \\
		& +\alpha\left[h\wedge\chi\right]+\left[h\wedge\tau\right]+\left[\omega\wedge\chi\right] \\
		& =\text{d}_\omega\chi+\alpha\left[h\wedge\chi\right]+\sigma\mu(1+\sigma\alpha)\left[e\wedge\chi\right]-\sigma\mu(1+\sigma\alpha)\left[\rho\wedge h\right] \\
		& -\Lambda_0\left[\rho\wedge e\right]+\left[h\wedge\tau\right],
	\end{aligned}
\end{align}
where we have defined three field-independent parameters  $\tau\equiv  i_\xi\omega, \rho\equiv i_\xi e$ and $\chi\equiv i_\xi h$. The above results suggest three distinct transformations: the internal Lorentz transformation ${\delta_\tau^{\text{j}}}$, the translation ${\delta_\rho^{\text t}}$, and a novel transformation $\delta_\chi^{\text m}$. With these transformations, \eqref{eq:Le-e}-\eqref{eq:Le-h} can be rewritten into a unified form as ${\cal L}_{\xi}(\bullet)=(\delta^{\text j}_\tau+\delta_\rho^{\text t}+\delta_\chi^{\text m})(\bullet)$ with $\bullet=(e,\omega,h)$  such that one can directly read off
\begin{align}\label{gauge one}
	\delta^{\text j}_\tau e = \left[ {e,\tau } \right],\quad\delta^{\text t}_\rho e = {\text{d}_\omega }\rho  + \alpha\left[ {h,\rho } \right],\quad\delta^{\text m}_\chi e = \alpha \left[ {e,\chi } \right];
\end{align}
\begin{align}\label{gauge two}
	\delta ^{\text j}_\tau\omega  = {\text{d}_\omega }\tau,\quad \delta ^{\text t}_\rho\omega  = \mu (1 + \sigma \alpha )\left[ {h,\rho } \right],\quad\delta ^{\text m}_\chi\omega  = \mu (1 + \sigma \alpha )\left[ {e,\chi } \right];
\end{align}
and
\begin{align}
	\begin{aligned}\label{gauge three}
		&\delta^{\text j}_\tau h = \left[ {h,\tau } \right],\quad\delta^{\text t}_\rho h = \sigma \mu (1 + \sigma \alpha )\left[ {h,\rho } \right] + {\Lambda _0}\left[ {e,\rho } \right],\\
		&\delta^{\text m}_\chi h = {\text{d}_\omega }\chi  + \alpha \left[ {h,\chi } \right] + \sigma \mu (1 + \sigma \alpha )\left[ {e,\chi } \right],
	\end{aligned}
\end{align}
respectively.
Remarkably, comparing against the usual theory in \cite{Achucarro:1986uwr, Geiller:2020edh, Kiran:2014dfa}, a new transformation $\delta_\varphi^{\text m}$ emerges in $\rm{MMG}$, which originates from the new degree of freedom $h$ and its related  equation of motion. By carefully comparing the three transformations operating on the fields in \eqref{gauge one}-\eqref{gauge three}, we conclude the following novel properties. (i) \eqref{gauge one} shows that the transformations $\delta^{\text m}_\chi$ and ${\delta_\tau^{\text j}}$ on $e$ differs only by a parameter $\alpha$. (ii) The Lorentz transformation ${\delta_\tau^{\text j}}$ on $e$ and $h$ has the similar structure. (iii) The translation transformation $\delta^{\text t}_\rho$ on $e$ includes the term $\alpha\left[ {h,\rho } \right]$ while the novel translation $\delta_\chi^{\text m}$ on $h$ includes both $\sigma \mu (1 + \sigma \alpha )\left[ {e,\chi } \right]$ and an intrinsic term $\alpha \left[ {h,\chi } \right]$, implying that  when $1+\sigma\alpha=0$, $\delta_{\chi}^{\text m}$ on $h$ has the same formula as $\delta_\rho^{\text t}$ on $e$ such that \eqref{gauge one} and \eqref{gauge three} coincides by switching $\rho$ and $\chi$.

We shall move on to verify that the transformations ${\delta ^{\text j}_\tau}$, ${\delta ^{\text t}_\rho}$ and ${\delta ^{\text m}_\chi}$ indeed generate symmetries, that's to say, their operations on Lagrangian $ L$ yield the total derivatives for arbitrary $f$, i.e., \eqref{definition of tansformation}. Their operations on the curvature are
\begin{align}\label{gauge four}
	{\delta ^{\text j}_\tau}R(\omega) = \left[ {R(\omega),\tau } \right],\quad{\delta ^{\text t}_\rho}R(\omega) = \mu (1 + \sigma \alpha ){\text{d}_\omega }\left[ {h,\rho } \right],\quad{\delta ^{\text m}_\chi}R(\omega) = \mu (1 + \sigma \alpha ){\text{d}_\omega }\left[ {e,\chi } \right],
\end{align}
and on the torsion are
\begin{align}\label{gauge five}
	\begin{aligned}
		&{\delta ^{\text j}_\tau}T(e) = \left[ {T(e),\tau } \right],\quad {\delta ^{\text t}_\rho}T(e) = \left[ {R(\omega ),\rho } \right] + \alpha {\text{d}_\omega }\left[ {h \wedge \rho } \right] + \mu (1 + \sigma \alpha )\left[ {\left[ {h,\rho } \right] \wedge e} \right], \\
		&{\delta ^{\text m}_\chi}T(e) = \alpha {\text{d}_\omega }\left[ {e,\chi } \right] + \mu (1 + \sigma \alpha )\left[ {\left[ {e,\chi } \right] \wedge e} \right].
	\end{aligned}
\end{align}
Their detailed  operations on the Lagrangian $L$ are shown in Appendix \ref{sec:B-1}, giving that
\begin{subequations}\label{action on L}
	\begin{align}
		\delta _\varphi ^{\text j}L &= \text{d}(I_\varphi ^{\text j}\theta  + {B_\varphi^1 }),\\
		\delta _\varphi ^{\text t}L &= \text{d}(I_\varphi ^{\text t}\theta  + {B_\varphi^2 }),\\
		\delta _\varphi ^{\text m}L &= \text{d}(I_\varphi ^{\text m}\theta  + {B_\varphi^3 }),
	\end{align}
\end{subequations}
in which we choose a general zero-form $\varphi$ to generate the transformations. These mean that the gauge symmetries can indeed be generated through these transformations without using  equations of motion. In the calculations,  we introduce three contractions $I_\varphi^{\text j}$, $I_\varphi^{\text t}$ and $I_\varphi^{\text m}$ generated by $\varphi$
\begin{subequations}\label{internal symmetris}
	\begin{align}
		&	I_\varphi ^{\text j}\theta  = {\text{d}_\omega }\varphi  \wedge ( - \sigma e + \frac{1}{2\mu}\omega ) + [e,\varphi ] \wedge h\label{contraction one},\\
		&	I_\varphi ^{\text t}\theta  = \mu(1 + \sigma \alpha )[h,\varphi ] \wedge ( - \sigma e + \frac{1}{2\mu}\omega ) + ({\text{d}_\omega }\varphi + \alpha \left[ {h,\varphi} \right]) \wedge h\label{contraction two},\\
		&I_\varphi ^{\text m}\theta  = \mu(1 + \sigma \alpha )[e,\varphi ] \wedge ( - \sigma e + \frac{1}{2\mu}\omega ) + \alpha [e,\varphi ] \wedge h,\label{contraction three}
	\end{align}
\end{subequations}
and
\begin{subequations}\label{Bs}
	\begin{align}
&B_\varphi ^1 = \frac{1}{\mu}{E_\omega } \wedge \varphi,\label{2.26a} \\
&B_\varphi ^2 = \frac{\alpha }{2}\left[ {h \wedge h} \right] \wedge \varphi  + \frac{{{\Lambda _0}}}{2}\left[ {e \wedge e} \right] \wedge \varphi  + h \wedge {{\rm{d}}_\omega }\varphi, \\
&B_\varphi ^3 = {E_h} \wedge \varphi  = 0.
	\end{align}
\end{subequations}
which can be  derived from \eqref{eq:total delta L} for the contractions $I_\varphi^{\text{j}} \delta e=\delta_\varphi^{\text{j}} e$, $I_\varphi^{\text{t}} \delta e=\delta_\varphi^{\text{t}} e$ and $I_\varphi^{\text{m}} \delta e=\delta_\varphi^{\text{m}} e$ (same to $\omega$ and $h$).

For the subsequent algebraic construction, we verify that the transformations form a closed algebra, which ensures the completeness of the symmetries.  This requires  that the anti-commutation relations of $\delta_\tau^{\text{j}}$, $\delta_\rho^{\text{t}}$, and $\delta_\chi^{\text{m}}$ can still be expressed via their operations. For example,  let us consider the double operations of $\delta _{{\varphi }}^{\text{t}}$ on $e$
	\begin{align}
		&\delta _{{\varphi _1}}^{\text{t}}\delta _{{\varphi _2}}^{\text{t}}e = \mu {(1 + \alpha \sigma )^2}\left[ {\left[ {h,{\varphi _1}} \right],{\varphi _2}} \right] + \alpha {\Lambda _0}\left[ {\left[ {e,{\varphi _1}} \right],{\varphi _2}} \right]\label{c.61},
	\end{align}
where the field-dependent terms (e.g., $\delta_{\delta_{\varphi_1}^{\text{t}} \varphi_2}^{\text{t}} e$ on the right-hand side of \eqref{c.61}) are omitted for clarity\footnote{Since the field-dependent components of the anti-commutative relations are mainly determined by the choice of $\varphi$. Consequently, here we solely focus on the algebra of the other fields beyond $\varphi$.}. Above expression  indicates that the transformation $\delta_\rho^{\text{t}}$ cannot form a closed algebra on $e$, because we cannot apply an operation to $e$ such that the result contains $h$ without the $\text{d}_\omega \varphi$ term, which can be explicitly seen from \eqref{gauge one}.

To proceed, we  propose a mapping from the three-dimensional space  $( \omega, e,h)$ to space $(\tilde \omega,\tilde e,h)$ via
	\begin{align}\label{larger space}	
		\tilde{\omega}=a_1e+b_1\omega+c_1 h,~~
		\tilde{e}=a_2e+b_2\omega+c_2h.
	\end{align}
Then we impose the constraints on the parameters in \eqref{larger space} by requiring consistency on $(\tilde{e}, \tilde{\omega})$
\begin{subequations} \label{eqalgebra:total}
	\begin{align}
&\delta_{\tau}^{\text{j}}\tilde{\omega}=d_{\tilde{\omega}}\tau\label{eqalgebra:sub1},  \\
		& \delta_\rho^{\text{t}}\tilde{\omega}=a_3[\tilde{e},\rho] \label{eqalgebra:sub2}, \\
		& \delta_{\tau}^{\text{j}}\tilde{e}=[\tilde{e},\tau] \label{eqalgebra:sub3}, \\
		& \delta_{\rho}^{\text{t}}\tilde{e}=a_{2}d_{\tilde{\omega}}\rho+b_{3}[\tilde{e},\rho],\label{eqalgebra:sub4}
	\end{align}
\end{subequations}
from which we can obtain
\begin{align}
\begin{aligned}\label{b.8constraints}
   &b_1 = 1,~~a_1=b_2=0,~~ a_3 a_2 = c_1 \Lambda_0 = \mu (1 + \sigma \alpha )\left( {1 + {c_1}\mu \sigma (1 + \sigma \alpha )} \right),\\
   &{a_2}b_3 = c_2\Lambda_0,~~{a_2}({c_1} - \alpha ) + {b_3}{c_2} = \mu (1 + \sigma \alpha )({c_2}\sigma  + {b_2}).
   \end{aligned}
\end{align}
Here we require that $a_2$ and $a_3$ are non-zero so that the parametric point $1+\alpha\sigma=0$ is inadmissible.  Subsequently, it is straightforward  to check that the anti-commutative operations $\delta_{\varphi_1}^{\text{j}}\delta_{\varphi_2}^{\text{j}}-\delta_{\varphi_2}^{\text{j}}\delta_{\varphi_1}^{\text{j}}$, $\delta_{\varphi_1}^{\text{t}}\delta_{\varphi_2}^{\text{t}}-\delta_{\varphi_2}^{\text{t}}\delta_{\varphi_1}^{\text{t}}$ and $\delta_{\varphi_1}^{\text{j}}\delta_{\varphi_2}^{\text{t}}-\delta_{\varphi_2}^{\text{t}}\delta_{\varphi_1}^{\text{j}}$  satisfy the closed algebras on space $(\tilde \omega, \tilde e)$
	\begin{align}\label{operator algebra two}
    \boxed{
		\left[ {\delta _{{\tau _1}}^{\text{j}},\delta _{{\tau _2}}^{\text{j}}} \right] = \delta _{\left[ {{\tau _1},{\tau _2}} \right]}^{\text{j}},\quad\left[ {\delta _\tau ^{\text{j}},\delta _\rho ^{\text{t}}} \right] = \delta _{\left[ {\tau ,\rho } \right]}^{\text{t}},\quad\left[ {\delta _{{\rho _1}}^{\text{t}},\delta _{{\rho _2}}^{\text{t}}} \right] = {{c_1}{\Lambda _0}}\delta _{\left[ {{\rho _1},{\rho _2}} \right]}^{\text{j}} + \frac{c_2\Lambda_0}{a_2}\delta _{\left[ {{\rho _1},{\rho _2}} \right]}^{\text{t}}.
        }
	\end{align}
These brackets resemble those in MB model \cite{Geiller:2020edh,Banerjee:2012jn,Banerjee:2009vf,Banerjee:2011cu}, which generates a $\mathfrak{g}_{\Lambda}$ algebra \cite{Cacciatori:2005wz,Giacomini:2006dr}. According to Appendix \ref{sec:B-2}, the closed algebras discussed above are not valid on field space $(\tilde \omega, \tilde e, h)$, as  one cannot apply the  rules \eqref{eqalgebra:total} on $h$. We also claim that the translation symmetry exhibits \textit{approximate completeness} at the limit $1+\sigma\alpha\to 0$. Additionally,  $\delta _{{\varphi}}^{\text{m}}$ is not included in the closed brackets \eqref{operator algebra two} and  we cannot simultaneously expand the closed algebra combining $\delta^{\text{t}}_\varphi$ and $\delta^{\text{m}}_\varphi$ on the space $(\tilde{\omega}, \tilde{e})$.
However, as the further derivations in Appendix \ref{sec:B-3} show, transition to an enlarged space $(\hat\omega, \tilde e ,h)$ at $1+\alpha\sigma=0$ admits closed algebras
\begin{align}\label{m algebras}
  \boxed{
   \left[ {\delta _{{\varphi _1}}^{\text{j}},\delta _{{\varphi _2}}^{\text{j}}} \right] = \delta _{\left[ {{\varphi _1},{\varphi _2}} \right]}^{\text{j}},\quad \left[ {\delta _{{\varphi _1}}^{\text{j}},\delta _{{\varphi _2}}^{\text{m}}} \right] = \delta _{[{\varphi _1},{\varphi _2}]}^{\text{m}},\quad \left[ {\delta _{{\varphi _1}}^{\text{m}},\delta _{{\varphi _2}}^{\text{m}}} \right] = \alpha \delta _{\left[ {{\varphi _1},{\varphi _2}} \right]}^{\text{m}}.
  }
\end{align}
We now summarize our studies in the closed algebras formed by the three transformations : (i) the internal Lorentz transformation  $\delta_\varphi^{\text{j}}$ and the translation transformation $\delta_\varphi^{\text{t}}$ form closed algebras in $(\tilde \omega, \tilde e)$ at general points of parameters; (ii) $\delta_\varphi^{\text{j}}$ and novel translation $\delta_\varphi^{\text{m}}$ form closed algebras in $(\hat \omega, \tilde e, h)$ at specific point $1+\sigma\alpha=0$. It is noted that the strategy  of changing the field space is actually to make sure the gauge symmetries have closed algebras. In general,  it is difficult for both the gauge symmetry and the algebra closure to simultaneously have closed properties, because one can't  freely add additional freedom to the system, especially in high-dimensional cases \cite{Zanelli:2005sa, Zanelli:2012px}. But as we will see in Section  \ref{fg one}, we can concentrate on the approximately closed properties of algebra \eqref{m algebras} in maximally symmetric solution of MMG.

\section{Gauge charges and diffeomorphism charge}
\label{section 3}
        	
In this section, we will use the method of covariant phase space to construct the gauge charges, arising from the three gauge symmetries introduced in the previous section.   This formalism facilitates the separation of the diffeomorphism charge into integrable and non-integrable components, which correspond to the  integrable charge and flux, respectively \cite{Barnich:2010eb, Compere:2018ylh, Compere:2020lrt}.  	

With the symplectic potential current $\theta$ of MMG given by \eqref{symplectic potential}, we now proceed to use the covariant phase space method. The variation of $\theta$ determines the symplectic structure
\begin{align}\label{symplectic form}
	\bar \Omega (\delta ,\delta ) = \int\limits_\Sigma  {\delta \omega  \wedge (\sigma \delta e - \frac{1}{{2\mu }}\delta \omega ) - \delta e \wedge \delta h},
\end{align}
where $\Sigma$ is a co-dimension one surface.  Then using the contractions  $I_\tau^{\text{j}}, I_\rho^{\text{t}}$ and $I_\chi^{\text{m}}$ (see Table \eqref{operators}) to contract with $\bar \Omega$ , we obtain three gauge charges
\begin{subequations}\label{gauge charges}
	\begin{align}
		&	-I_\tau^{\text{j}} \bar\Omega=  {\slashed{\delta}} {{Q_\tau ^{\text{j}}} }=\int\limits_{\cal S} {{\slashed{\delta}}q_\tau ^{\text{j}}} \approx \int\limits_{\cal S} {\tau \delta (\frac{1}{\mu }\omega- \sigma e  )},\label{gauge charge 1}\\
		&	-I_\rho^{\text{t}} \bar\Omega	 = {\slashed{\delta}} {{Q_\rho ^{\text{t}}} }=\int\limits_{\cal S} {{\slashed{\delta}}q_\rho ^{\text{t}}}\approx \int\limits_S {\rho \delta ( h-\sigma \omega  )}, \\
		&	-I_\chi^{\text{m}} \bar\Omega=  {\slashed{\delta}} {{Q_\chi ^{\text{m}}} }=\int\limits_{\cal S} {{\slashed{\delta}}q_\chi ^{\text{m}}}  \approx \int\limits_{\cal S} {\chi \delta e},\label{gauge charge 3}
	\end{align}
\end{subequations}
where $\slashed{\delta}$ means that these charges have no need to be integrable. We name the charge $Q_\chi ^{\text{m}}$ as auxiliary charge due to the field $h$.

The three transformations defined in \eqref{gauge one}-\eqref{gauge three} are on-shell equivalent with the Lie derivative ${\cal L}_\xi$ as addressed in the previous section. Now let us contract $\bar\Omega (\delta,\delta )$ along the direction of diffeomorphism to extract the diffeomorphism charge $D_\xi$
\begin{align}
	\begin{aligned}\label{diffeomorphism charge}
		& \bar\Omega ({{\cal L}_\xi },\delta ) = \slashed{\delta} D_\xi\\
    &= \int\limits_\Sigma  {\left( {{\text{d}_\omega }{i_\xi }h + {i_\xi }{\text{d}_\omega }h + [h,{i_\xi }\omega ] - \sigma ({\text{d}_\omega }{i_\xi }\omega  + {i_\xi }R(\omega ))} \right) \wedge \delta e}+ \left( {{\text{d}_\omega }{i_\xi }e + {i_\xi }T(e)} \right.+\left. {  [e,{i_\xi }\omega ]} \right) \wedge \delta h  \\
		&~~+ \left( {\frac{1}{\mu }({\text{d}_\omega }{i_\xi }\omega  - \sigma \mu {\text{d}_\omega }{i_\xi }e) - \sigma ({i_\xi }T(e) + \frac{1}{\mu }{i_\xi }R(\omega ) + [e,{i_\xi }\omega ])} \right) \wedge \delta \omega \\
    & = \int\limits_\Sigma  {{i_\xi }\left( {({\text{d}_\omega }h - \sigma R(\omega ) + \frac{\alpha }{2}\left[ {h \wedge h} \right]) \wedge \delta e + (T(e) + \alpha \left[ {e \wedge h} \right]) \wedge \delta h} \right.} \\
		&~~\left. { + (\sigma T(e) - \frac{1}{\mu }R(\omega ) - \left[ {e \wedge h} \right]) \wedge \delta \omega } \right) + \delta \left( {{i_\xi }e \wedge (\sigma R(\omega ) - {\text{d}_\omega }h - \frac{{{\Lambda _0}}}{2}\left[ {e \wedge e} \right]) - \frac{\alpha }{2}\left[ {h \wedge h} \right])} \right.\\
		&~~- {i_\xi }h \wedge (T + \alpha \left[ {e \wedge h} \right])\left. { + {i_\xi }\omega  \wedge (\frac{1}{\mu }R(\omega ) - \sigma T(e) + \left[ {e \wedge h} \right])} \right) + \int\limits_{\cal S} {\left( {{i_\xi }h \wedge \delta e + {i_\xi }e \wedge \delta h } \right.}  \\
		&~~ - \sigma {i_\xi }\omega  \wedge \delta e- \frac{1}{\mu }{i_\xi }\omega  \wedge \delta \omega \left. { - \sigma {i_\xi }e \wedge \delta \omega } \right)
	\end{aligned}
\end{align}
 with an arbitrary field-independent vector $\xi$, i.e., $\delta\xi=0$. In the second equality of \eqref{diffeomorphism charge}, we recall the identities ${{\cal L}_\xi }\omega  = {\text{d}_\omega }{i_\xi }\omega  + {i_\xi }R(\omega )$ , ${{\cal L}_\xi }e = {\text{d}_\omega }{i_\xi }e + {i_\xi }T(e) + [e,{i_\xi }\omega ]$, ${{\cal L}_\xi }h = {\text{d}_\omega }{i_\xi }h + {i_\xi }{\text{d}_\omega }h + [h,{i_\xi }\omega ]$ and invoke relation \eqref{domega on fields}, while in the third equality we consider the vector $\xi$ to be not only tangent to $\Sigma$, but also to $\cal S$ such as to add term $\frac{\Lambda }{2}{i_\xi }(\delta e \wedge [e \wedge e]) = \frac{{{\Lambda _0}}}{2}\delta ({i_\xi }e \wedge [e \wedge e])$.

When the equations of motion \eqref{EOM} are satisfied, it is obvious that the diffeomorphism charge $\slashed{\delta} D_{\xi}$ can be reduced into
\begin{align}\label{difeo charge}
	\slashed{\delta} {{D_\xi} }=\int\limits_\Sigma  {\delta {q_\xi } - {i_\xi\theta }},
\end{align}
with the integrable charge
\begin{align}
	\begin{aligned}
		& q_{\xi}=i_{\xi} \omega \wedge\left(-\sigma e+\frac{1}{2 \mu} \omega\right)+i_{\xi} e \wedge h \label{integrable charge},
	\end{aligned}
\end{align}
and flux $i_\xi \theta$.
In addition, adopting  the method in \cite{Jacobson:2015uqa}, we can obtain the Noether charge $Q_\xi$  of $\Sigma$ relative to $\xi$  in MMG via
\begin{align}\label{Integrable charge}
	\text{d}Q_\xi=\theta({\cal L}_\xi)-i_\xi L,
\end{align}
where $\theta({\cal L}_\xi)$ and $i_\xi L$ are calculated by
\begin{subequations}\label{thetaq}
	\begin{align}
		&\theta ({{\cal L}_\xi }) = {{\cal L}_\xi }\omega  \wedge ( - \sigma e + \frac{1}{{2\mu }}\omega ) + {{\cal L}_\xi }e \wedge h\notag\\
		&= \text{d}({i_\xi }\omega  \wedge \frac{{\partial L}}{{\partial {\text{d}_\omega }\omega }} + {i_\xi }e \wedge \frac{{\partial L}}{{\partial T(e)}}) - {i_\xi }\omega  \wedge {\text{d}_\omega }\frac{{\partial L}}{{\partial {\text{d}_\omega }\omega }} + {i_\xi }e \wedge {\text{d}_\omega }\frac{{\partial L}}{{\partial T(e)}}\notag\\
		&~~~~+ {i_\xi }R(\omega ) \wedge \frac{{\partial L}}{{\partial {\text{d}_\omega }\omega }} + {i_\xi }T(e) \wedge \frac{{\partial L}}{{\partial T(e)}},\label{itheta2}\\
		{i_\xi }L &= {i_\xi }R(\omega ) \wedge \frac{{\partial L}}{{\partial {\text{d}_\omega }\omega }} + {i_\xi }T(e) \wedge \frac{{\partial L}}{{\partial T(e)}} + {i_\xi }\omega  \wedge \frac{{\partial L}}{{\partial \omega }} + {i_\xi }e \wedge \frac{{\partial L}}{{\partial e}} + {i_\xi }h \wedge \frac{{\partial L}}{{\partial h}}.
	\end{align}
\end{subequations}
Here the results of ${\cal L}_\xi \omega$, ${\cal L}_\xi e$ in \eqref{domega on fields} are used again, and one can check that the Noether charge $Q_\xi$ in the total differential term in \eqref{itheta2} accords with $\text{d}q_\xi$ from \eqref{integrable charge}, which are both consistent with the results in \cite{Setare:2015pva} for $\rm{MMG}$.

Combining  the expressions \eqref{gauge charges} and \eqref{difeo charge}, we obtain a relation  between the three gauge charges and diffeomorphism charge
 \begin{align}\label{suming}
	\slashed{\delta} {{D_\xi}} \approx\slashed{\delta} {{Q_\tau ^{\text{j}}} }+\slashed{\delta} {{Q_\rho ^{\text{t}}} }+\slashed{\delta} {{Q_\chi ^{\text{m}}} }.
\end{align}
This result is reasonable because the definitions of the three transformations actually derive from the equivalence  between the Lie derivative and $\delta_\tau^{\text{j}}, \delta_\rho^{\text{t}}$ and $\delta_\chi^{\text{m}}$ up to some on-shell vanishing terms.

\section{New exact symmetries in MMG}
\label{section 4}
An exact symmetry that preserves the symplectic structure corresponds to a Hamiltonian vector field on phase space.\footnote{Here, we consider field-dependent transformations, denoted by $\delta_\xi$. The details of this field-dependent variation will be further explained later.} In Wald's method, the conserved charge associated with this symmetry is the Hamiltonian that generates the corresponding flow on symplectic manifold \cite{Lee:1990nz, Harlow:2019yfa}. Specifically, in the presence of a Killing vector $\xi_K$, the exact symmetries for the field $\Phi$ are generated by $(\xi_K,\varphi(\xi_K))$, satisfying
\begin{align}
\delta_{(\xi_K,\varphi(\xi_K))}\Phi\approx0,
\end{align}
which implies that the gauge transformations keep the solution space invariant. Consequently, the associated exact symmetries correspond to the on-shell conserved charges $k_\xi$ \cite{Frodden:2019ylc}, i.e.,
\begin{align}\label{conserved prperty}
\text d k_{\xi_K} \approx  0,
\end{align}
where $k_\xi$ is obtained by the contraction of the exact symmetries on the symplectic current.

In the triad formalism, the exact symmetries are usually generated by Kosmann derivative (also known as Lie-Lorentzian derivative)  \begin{eqnarray}\label{eq=Kasmann}
{{\cal K}_\xi }: = {{\cal L}_\xi } + \delta _{\lambda_\xi }^{\text{j}} ~~~\mathrm{with}~~~  \lambda_\xi^I= - \frac{1}{2}{\varepsilon ^I}_{JK}{e^{\mu J}}{{\cal L}_\xi }{e_\mu }^K,
\end{eqnarray}
which was introduced to remedy the defect when ${\cal L}_\xi e\ne0$ to adapt the Wald's method \cite{Jacobson:2015uqa, DePaoli:2018erh, Prabhu:2015vua,Geiller:2020edh,JACKIW1980257, Obukhov:2006ge, fatibene2009general}.

In MMG, we find that both Lie and Kosmann derivative closely fail to generate exact symmetries,  while also forming the closed translational algebra  \eqref{operator algebra two} in the limit $1+\sigma\alpha\to0$ and the transformations involving $\delta_\varphi^{\text{m}}$. To address this issue, we  shall propose a new transformation constructed from $\delta_\varphi^{\text{j}}$ and $\delta_\varphi^{\text{m}}$. We then deduce the resulting exact symmetry and its conserved charge using Wald's method. We conclude this section by applying this new transformation to the rotational BTZ black hole solution in MMG.

 \subsection{Transformations generating exact symmetries and their corresponding conserved charges}
  In $\rm{MMG}$, there are three fields $e,\omega$ and $h$, and we  expect to find a transformation $\delta_{(\xi,\varphi(\xi))}$,  which has the following  properties: (i) It can generate gauge symmetries fulfilling  \eqref{definition of tansformation}; (ii) A Killing vector $\xi_K$ shall give
\begin{align}\label{condition of variations0}
  \delta_{({\xi_{K},\varphi(\xi_{K}))}}e\approx 0,\quad  \delta_{(\xi_{K},\varphi(\xi_{K}))}\omega\approx 0,\quad  \delta_{(\xi_{K},\varphi(\xi_{K}))}h\approx 0.
\end{align}
When considering only the Lorentz transformation, both ${\cal L}_\xi$ and ${\cal K}_\xi$ acting on $e$, $\omega$ and $h$ possess the above properties, allowing us to Wald's method to study the conserved charge. However, when we consider the new metric-preserving symmetry generated by $\delta_\varphi^{\text{m}}$, we find that neither ${\cal L}_\xi$ nor ${\cal K}_\xi$ retains this feature, rendering Wald's method invalid. For instance, a transformed field $e'=e+\delta _\varphi ^{\rm{m}}e$ in general does not  satisfy ${\cal L}_{\xi_K}e'\approx0$, and the same holds for $\omega'$ and $h'$. Therefore, an issue naturally arises: \textit{For the latter, is it possible to construct the  transformations $\delta_{(\xi,\varphi(\xi))}$ containing the transformation $\delta _\varphi ^{\rm{m}}$ to allow us to use Wald's method? }

The answer is yes and following is our proposal.  We adopt a linear combination of $\delta_\varphi^\text{j}$ and $\delta_\varphi^\text{m}$ to construct ${\delta _{(\xi ,\varphi (\xi ))}}$ as
\begin{align}\label{new variation}
  \boxed{
    \delta_{(\xi, \varphi(\xi))} = \delta_{(\xi, \beta_\xi, \bar{\beta}_\xi)} := \mathcal{L}_\xi + \delta_{\beta_\xi}^{\mathrm{j}} + \delta_{\bar{\beta}_\xi}^{\mathrm{m}},
  }
\end{align}
where the two Lorentz vectors ${\beta_\xi }$ and ${\bar\beta _\xi }$ are to be determined and should  satisfy\footnote{It is noted that \eqref{51a} does not strictly require $\xi$ to be a Killing vector field, however,  \eqref{51b1} and \eqref{51b} are required to be held when $\xi$ is a Killing vector, which relaxes the requirement for the existence of $\beta_\xi$. }
\begin{subequations}\label{condition on new variation}
\begin{empheq}[box=\boxed]{align}
  &\beta_\xi + \alpha\bar{\beta}_\xi = \lambda_\xi, \label{51a} \\
  &\delta_{(\xi_K, \beta_{\xi_K}, \bar{\beta}_{\xi_K})} E_h \approx 0, \label{51b1} \\
  &\mathcal{L}_{\xi_K} h' + \mathrm{d}_\omega \bar{\beta}_{\xi_K} + [h', \lambda_{\xi_K}] + \sigma \mu (1 + \sigma \alpha) [e', \bar{\beta}_{\xi_K}] \approx 0. \label{51b}
\end{empheq}
\end{subequations}

Subsequently, it is straightforward to derive
\begin{align}\label{condition of variations}
   {\delta _{({\xi_K} ,{\beta _{\xi_K} },{\bar \beta _{\xi_K} })}}e'\approx 0,\quad   {\delta _{({\xi_K} ,{\beta _{\xi_K} },{\bar \beta _{\xi_K} })}}\omega'\approx 0,\quad   {\delta _{({\xi_K} ,{\beta _{\xi_K} },{\bar \beta _{\xi_K} })}}h'\approx 0.
\end{align}

In order to give the concrete expression of ${\delta _{(\xi ,{\beta _\xi },{\bar \beta _\xi })}}$, we must solve out the formulas of ${\beta _\xi }$ and ${\bar\beta _\xi }$. It is noted that as addressed in \cite{Jacobson:2015uqa}, when ${\delta _{(\xi ,\varphi(\xi))}}$ is the Kosmann derivative excluding the new symmetry in MMG,  the condition \eqref{51a} and \eqref{51b1} are sufficient to determine ${\beta _\xi }$ and ${\bar\beta _\xi }$. However,  when the new symmetry is considered, we find that even the condition of \eqref{51a}  accompany with other assumptions are insufficient to solve the problem and the details are present in Appendix \ref{appendix c1}.  This motivates us to introduce another condition which we propose to be \eqref{51b}, the deriving of which is shown in Appendxi \ref{appendix c2}. Specifically, we will see that to satisfy condition \eqref{51b}, The parameter combination $1+\sigma\alpha$ can not be arbitrary, and its order must be consistent with the order of the infinitesimal gauge transformations which have approximately closed algebras.

In addition,  since \eqref{51b} is  a first-order equation of $\bar \beta_\xi$, its solution should satisfy the integrable condition (or it should be compatible with the equations of motion) due to the relation ${{\rm{d}}_\omega }{{\rm{d}}_\omega }{{\bar \beta }_{{\xi _K}}} = \left[ {R(\omega ),{{\bar \beta }_{{\xi _K}}}} \right]$.  According to Appendix \ref{appendix c2} ,   $\bar \beta_\xi$ can be compatible with the following integrable condition
\begin{align}\label{parameters constraints}
{\Lambda _0} = \frac{{1 - {\ell ^2}{\sigma ^2}{\mu ^2}{{(1 + \sigma \alpha )}^2}}}{{\alpha {\ell ^2}}}.
    \end{align}
 Note that $1+\sigma\alpha=0$ is not allowed here.

Consequently, we can determine  $\beta_\xi$ and $\bar \beta_\xi$ by using  \eqref{condition on new variation}, and further obtain the corresponding  transformation $\delta_{(\xi,\varphi(\xi))}$ which fulfills the conditions \eqref{condition of variations} for the three fields. In this framework, the Wald's method becomes valid in transformations related $\delta_\varphi^{\text m}$.
Subsequently, using the method regarding to the exact symmetries \cite{Wald:1993nt, Iyer:1994ys, Ghodrati:2016vvf, Kastor:2009wy, Xiao:2023lap, Hajian:2015xlp}, we can derive the conserved charges by the contraction of symplectic structure \eqref{diffeomorphism charge}  as
\begin{align}\label{exact diffeomorphism charge}
    \bar \Omega ({\delta _{({\xi _K},\beta ({\xi _K}),\bar \beta ({\xi _K}))}},\delta ) \approx \slashed{\delta} {D_{{\xi _K}}} + \delta Q_{\beta ({\xi _K})}^{\rm{j}} + \delta Q_{\bar \beta ({\xi _K})}^{\rm{m}}\approx0.
\end{align}
Here, the non-integrable part is ambiguous.
By referring to the method proposed in \cite{Geiller:2020edh},  we can adopt the Dirichlet boundary conditions on AdS boundary to fix it as an integrable charge $\tilde D_{\xi}$. Note that this method has also been used in other theories, see for examples \cite{Kraus:2005zm,Miskovic:2006tm, Grumiller:2015xaa}. Thus,
the conserved charge is
\begin{align}\label{new charge}
  \boxed{
    D_{(\xi ,{\beta _\xi },{{\bar \beta }_\xi })}^N: = {\tilde D_\xi } + Q_{{\beta _\xi }}^{\rm{j}} + Q_{{{\bar \beta }_\xi }}^{\rm{m}},
  }
\end{align}
where $Q_{{\beta _\xi }}^{\rm{j}}$ and $Q_{{{\bar \beta }_\xi }}^{\rm{m}}$ can be obtained from \eqref{gauge charges}.

According to \cite{Wald:1993nt, Iyer:1994ys}, integrating the conserved charge \eqref{new charge} over the bifurcation surface ${\cal B}_H$ and the asymptotic boundary ensures that the first law of black holes, including the auxiliary field $h$, holds.

To summarize, we have proposed the conditions \eqref{condition on new variation} and a new transformation \eqref{new variation}, which generate a complete exact symmetry associated with a Killing vector in MMG, acting on the three relevant fields. This symmetry leads to the new conserved charge \eqref{new charge}, which includes the auxiliary charge $Q_{\bar \beta_\xi}^\text{m}$. In the next subsection, we will apply this proposal to the rotational BTZ black hole solution in MMG to determine the parameter regime where such an exact symmetry can be achieved.



\subsection{Example}\label{fg one}
$\rm{MMG}$ has maximally symmetric solution in the equation of motion \cite{MahdavianYekta:2015tmp}
\begin{align}\label{maxisolution}
R(\Omega ) =  - \frac{1}{{2{\ell ^2}}}\left[ {e \wedge e} \right],
\end{align}
whose solution of metric is the rotational BTZ black hole
\begin{align}\label{btz metric}
{\rm{d}}{s^2} =  - \frac{{({r^2} - {r_ + }^2)({r^2} - {r_ - }^2)}}{{{\ell ^2}{r^2}}}{\rm{d}}{t^2} + \frac{{{\ell ^2}{r^2}}}{{({r^2} - {r_ + }^2)({r^2} - {r_ - }^2)}}{\rm{d}}{r^2} + {r^2}{\left( {{\rm{d}}{\phi } + \frac{{{r_ + }{r_ - }}}{{\ell {r^2}}}{\rm{d}}t} \right)^2}.
\end{align}
The solution of $h$ is $h=C\mu e$ with $C=\frac{{1 - \alpha {\ell ^2}{\Lambda _0}}}{{2{\ell ^2}{\mu ^2}{{(1 + \sigma \alpha )}^2}}}$. Note that the maximally symmetric solution here is essentially different from that to the equation of motion \eqref{eoms} at the point $1+\sigma\alpha=0$. They belong to solutions of the same form but in different parameter regimes, because they have different closed algebra \eqref{operator algebra two} and \eqref{m algebras}, respectively.

With the internal metric $\eta^{IJ}$ (see Appendix \ref{appendix one}) and \eqref{btz metric}, we can read off the triad and the torsion free spin-connection $\Omega$
\begin{align}\label{hsolution}
{e_\mu }^I = \left( {\begin{array}{*{20}{c}}
{\frac{{\sqrt {({r^2} - r_ + ^2)({r^2} - r_ - ^2)} }}{{\ell r}}}&0&{  \frac{{{r_ + }{r_ - }}}{{\ell r}}}\\
0&{\frac{{\ell r}}{{\sqrt {({r^2} - r_ + ^2)({r^2} - r_ - ^2)} }}}&0\\
0&0&r
\end{array}} \right),\quad  {\Omega _\mu }^I = \left( {\begin{array}{*{20}{c}}
0&0&{\frac{r}{{{\ell ^2}}}}\\
0&{\frac{{{-r_ + }{r_ - }}}{{r\sqrt {({r^2} - r_ + ^2)({r^2} - r_ - ^2)} }}}&0\\
{\frac{{\sqrt {({r^2} - r_ + ^2)({r^2} - r_ - ^2)} }}{{\ell r}}}&0&{  \frac{{{r_ + }{r_ - }}}{{\ell r}}}
\end{array}} \right).
\end{align}
Since the conserved charges must satisfy the first law of the black hole, we will first derive this law for the BTZ black hole using the exact symmetries generated by the Lie and Kosmann derivatives. We will then verify that the new conserved charge \eqref{new charge}, obtained from the new transformation, can reproduce this result.

\begin{itemize}
\item \textbf{Exact symmetry of Lie derivative}.

 Since the rotational BTZ black hole \eqref{btz metric} has a Killing vector $ \xi_K=\left( {1,0,\frac{r_-}{\ell r_+}} \right)$\footnote{This Killing vector field vanishes on the bifurcation surface ${\cal B}_H$, thereby ensuring the integrability of the conserved charge, see \cite{MahdavianYekta:2016kqh}.}, so according to \eqref{second connection} and \eqref{hsolution}, the Lie derivatives of the three fields $e$, $\omega$ and $h$ are
\begin{align}\label{le3=0}
  {{\cal L}_{{\xi _K}}}{e_\mu}^I = 0,\quad {{\cal L}_{{\xi _K}}}{\omega_\mu}^I  = 0,\quad {{\cal L}_{{\xi _K}}} {{{  h}_\mu }^I}= 0.
\end{align}
Thus, $\lambda_\xi$ in \eqref{eq=Kasmann} vanishes such that the Lorentz vectors take $\beta_\xi=\bar \beta_\xi=0$. These imply that the corresponding new conserved charge \eqref{new charge} coincides with the diffeomorphism charge \eqref{difeo charge} associated with the Lie derivative, i.e., $D_\xi^N=D_\xi$.

Moreover, the integral in the charge $D_\xi$ can be splited into two parts: on the bifurcation surface ${\cal B}_H$ ($r\to r_+$)  and on $\rm{AdS}$ boundary ${ \Sigma}_{AdS}$ ($r\to\infty$), respectively. On $\rm{AdS}$ boundary, we add the Gibbon-Hawking term  \cite{Alishahiha:2015whv}
\begin{align}
    {S_{GH}} =  - \sigma \int\limits_{{{\Sigma}_{AdS} }} {\left[ {e \wedge \Omega } \right]} ,
\end{align}
and then follow the method proposed in \cite{Geiller:2020edh} to cancel  the non-integrable part in $i_\xi\theta$ with the boundary condition  ${\left. {(\theta  - \sigma \delta \left[ {e \wedge \Omega } \right])} \right|_{{{\cal \Sigma}_{\Sigma_{AdS}} }}} = 0$. Then the integral in  $D_\xi$ on the AdS boundary is
\begin{align}
  \int\limits_{{{\cal \Sigma}_{\Sigma_{AdS}} }} {\left( {{q_{{\xi _K}}} - \sigma {i_{{\xi _K}}}(e \wedge \Omega )} \right)}  = E + {\Omega _H}J,
\end{align}
where  ${\Omega _H} =   \frac{{{r_ - }}}{{{\ell r_+ }}}$ is the angular velocity and
\begin{align}\label{eq:E and J}
    E=\left( {(\sigma  + \alpha C)\frac{{r_ + ^2 + r_ - ^2}}{{2{\ell^2}}} + \frac{1}{{\mu \ell}}\frac{{{r_ + }{r_ - }}}{{{\ell^2}}}} \right),\quad J=\left( {(\sigma  + \alpha C)\frac{{{r_ + }{r_ - }}}{\ell} + \frac{1}{{\mu \ell}}\frac{{r_ + ^2 + r_ - ^2}}{{2\ell}}} \right)
\end{align}
are the energy and angular momentum, respectively.
The entropy is defined as (see example in \cite{Wei:2018sqy, Setare:2015pva})
\begin{align}\label{eq:S}
   \frac{{2\pi }}{{{\kappa _H}}}\int\limits_{{\cal B}_H} {{q_{{\xi _K}}}}  = {A_h}\left( {(\sigma  + \alpha C) + \frac{1}{{\mu \ell}}\frac{{{r_ - }}}{{{r_ + }}}} \right) = S,
\end{align}
where  $\kappa_H=\frac{{{r_ + }}}{{{\ell ^2}}}(1 - \frac{{{r_ - }^2}}{{{r_ + }^2}})$ is the surface gravity and $A_h=2\pi r_+$. Our expression of entropy reproduces the result obtained from \cite{Clement:2003sr, MahdavianYekta:2016kqh, Setare:2015pva}, and the energy and angular momentum also coincide with those from the canonical method in $\rm{MMG}$ \cite{MahdavianYekta:2015tmp}.

It is easy to check from \eqref{eq:E and J} and \eqref{eq:S} that
\begin{eqnarray}\label{eq:first law}
\text{d}E=2T_H \text{d}S+\Omega_H \text{d}J,
\end{eqnarray}
which is nothing but the modified form of differential first law of black hole \cite{Clement:2003sr,Moussa:2003fc,MahdavianYekta:2015tmp}.


\item\textbf{Exact symmetries of Lorentz transformation $\delta_\varphi^\text{j}$}.

We perform Lorentz transformation on triad $e$ and the torsion-free spin connection $\Omega$, and require that they preserve the metric at leading order. Specifically, we consider a Lorentz vector $\vartheta=\left( {{\vartheta _1}(t,\phi ),0,0} \right)$, which generates three new fields as
\begin{align}\label{newfields}
    &e^{new} = e + \delta _\vartheta ^{\rm{j}}e,\quad\omega ^{new} = \omega  + \delta _\vartheta ^{\rm{j}}\omega,\quad h^{new} = h + \delta _\vartheta ^{\rm{j}}h
\end{align}
and satisfies $\delta _\vartheta ^{\rm{j}} g_{\mu\nu}={\cal O}(\vartheta ^2)$, meaning that the transformation $\delta _\vartheta ^{\rm{j}}$ preserves the black hole solution \eqref{btz metric} or different triads correspond to the same metric. The Lie derivatives of the new fields along the Killing vector are
\begin{subequations}
    \begin{align}\label{new fields one}
&{{\cal L}_{{\xi _K}}}{e^{new}} = \frac{1}{{C\mu }}{{\cal L}_{{\xi _K}}}{h^{new}} =\left( {\begin{array}{*{20}{c}}
0&{\frac{{{r_ + }{r_ - }\varepsilon }}{{\ell r}}}&0\\
0&0&{ - \frac{{\ell r\varepsilon }}{{\sqrt {({r^2} - {r_ + }^2)({r^2} - {r_ + }^2)} }}}\\
0&{r\varepsilon }&0
\end{array}} \right),\\
&{{\cal L}_{{\xi _K}}}{\omega ^{new}} = {{\cal L}_{{\xi _K}}}{\Omega ^{new}} - \alpha {{\cal L}_{{\xi _K}}}{h^{new}} = \left( {\begin{array}{*{20}{c}}
{{\partial _t}\varepsilon }&{\frac{{r\varepsilon }}{{{\ell ^2}}}}&0\\
0&0&{\frac{{{r_ + }{r_ - }\varepsilon }}{{r\sqrt {({r^2} - r_ + ^2)({r^2} - r_ - ^2)} }}}\\
{{\partial _\phi }\varepsilon }&{\frac{{{r_ + }{r_ - }\varepsilon }}{{\ell r}}}&0
\end{array}} \right) - \alpha {{\cal L}_{{\xi _K}}}{h^{new}},
\end{align}
\end{subequations}
with ${\varepsilon} = \frac{{{r_ - }{\partial _\phi }{\vartheta _1}}}{{\ell {r_ + }}} + {\partial _t}{\vartheta _1}$. The condition of vanishing ${{\cal L}_\xi }e^{new}$, ${{\cal L}_\xi }\omega^{new}$ and ${{\cal L}_\xi }h^{new}$ is $\varepsilon=0$, which means that the gauge $\varepsilon\ne0$ breaks the existence of the exact symmetries related with the Lie derivative on the new fields. To cure this defect, one usually considers the Kosmann derivative instead of Lie derivative to apply in symplectic form, i.e.,
\begin{align}
    \bar \Omega ({{\cal K}_\xi }(\lambda_\xi),\delta ) = {D_\xi }(e,\omega,h) + Q_{\lambda_\xi}^{\rm{j}},
\end{align}
where $\lambda_\xi = \left( {\begin{array}{*{20}{c}}
{ - \varepsilon }&0&0
\end{array}} \right)$, ${D_\xi }(e,\omega,h)$ is expressed in  \eqref{difeo charge} and
\begin{align}\label{Kosmann charge}
    Q_{\lambda_\xi}^{\rm{j}}=\int\limits_{\cal S} {{\lambda _\xi }( - \sigma e + \frac{1}{{2\mu }}\omega )}
\end{align}
is calculated from \eqref{gauge charges},  which accords with one used in higher dimensional cases \cite{DePaoli:2018erh, Oliveri:2019gvm}. Then, with the same procedures in the previous case, it is straightforward to derive the first law of black hole \eqref{eq:first law}.
\item\textbf{Exact symmetries of new transformation including $\delta_\varphi ^\text{m}$}.

With the same fields $e^{new}$, $\omega^{new}$, $h^{new}$ but the Lorentz vector rescaled by $\vartheta  \to (1 + \sigma \alpha )\vartheta $ (or $\lambda_\xi\to(1 + \sigma \alpha )\lambda_\xi$),  we find that the three conditions in \eqref{condition on new variation} indeed can be satisfied in the linear order of the small quantity $(1+\sigma\alpha)$. The Lorentz vectors $\bar\beta_\xi$ and $ \beta_\xi$ are chosen as
\begin{subequations}
\begin{align}
&{{\bar \beta }_\xi } = \left( {{{\bar \beta }_1},{{\bar \beta }_2},{{\bar \beta }_3}} \right),\\
&{\beta _\xi } = \left( {-(1 + \sigma \alpha )\varepsilon  - \alpha {{\bar \beta }_1}, - \alpha {{\bar \beta }_2}, - \alpha {{\bar \beta }_3}} \right),
\end{align}
\end{subequations}
which satisfy $\beta_\xi+\alpha\bar\beta_\xi=\left( { - (1 + \sigma \alpha )\varepsilon ,0,0} \right)=\lambda_\xi$\footnote{Here, we relabel the form $(1+\alpha\sigma)\lambda_\xi$ as $\lambda_\xi$.}, then the transformation \eqref{new variation} gives
\begin{subequations}
    \begin{align}
&{\delta _{({\xi _K},{\beta _{{\xi _K}}},{{\bar \beta }_{{\xi _K})}}}}{e^{new}} \approx 0+{\cal O}\left( {{{(1 + \sigma \alpha )}^2}} \right),\\
&{\delta _{({\xi _K},{\beta _{{\xi _K}}},{{\bar \beta }_{{\xi _K})}}}}{E_h}({e^{new}},{\Omega ^{new}}) = {\delta _{({\xi _K},{\beta _{{\xi _K}}},{{\bar \beta }_{{\xi _K})}}}}{{\rm{d}}_{{\Omega ^{new}}}}{e^{new}} \approx 0+{\cal O}\left( {{{(1 + \sigma \alpha )}^2}} \right).\label{eq-66b}
    \end{align}
\end{subequations}
Note that \eqref{eq-66b} is equivalent with $ {\delta _{({\xi _K},{\beta _{{\xi _K}}},{{\bar \beta }_{{\xi _K})}}}}{\Omega ^{new}} \approx 0+{\cal O}\left( {{{(1 + \sigma \alpha )}^2}} \right)$.
To solve $\bar\beta_\xi$, we combine the condition \eqref{51b} or \eqref{first order euqation of bar beta}, which provides a first order equation for $\bar\beta_\xi$ as
\begin{align}\label{db}
\boxed{
    {\rm{d}}{{\bar \beta }_{{\xi _K}}} + \left[ {{\omega ^{new}} + \sigma \mu (1 + \sigma \alpha ){e^{new}},{{\bar \beta }_{{\xi _K}}}} \right] = 0.
    }
\end{align}
Under the integrable condition \eqref{parameters constraints}, the solution of ${{{\bar \beta }_{{\xi _K}}}}$ can vanish, or take the form in the path-ordered exponential $P\exp$\footnote{This fact arises because matrix ${{\omega ^{new}} + \sigma \mu (1 + \sigma \alpha ){e^{new}}}$ is generally non-commutative and the exact form of the integral is precisely the Dyson series.}:
\begin{align}\label{path order solution}
   {{\bar \beta }_{{\xi _K}}} = \left[ {P\exp \left( { - \int\limits_{\cal C} {{\omega ^{new}} + \sigma \mu (1 + \sigma \alpha ){e^{new}}} } \right) \wedge {\rm{d}}\psi } \right],
\end{align}
where $\cal C$ is a three-dimensional path and $\psi$ is an Lorentz vector satisfying
\begin{align}\label{boundary condition of one-form}
    \left[ {P\exp \left( { - \int\limits_{\cal C} {\omega  + \sigma \mu (1 + \sigma \alpha )e} } \right) \wedge {\rm{d}}\psi } \right] = 0.
\end{align}

Moreover, it is straightforward to check that the solutions of ${{\beta }_{{\xi _K}}}$ and ${{\bar \beta }_{{\xi _K}}}$ yield
\begin{align}
{\delta _{({\xi _K},{\beta _{{\xi _K}}},{{\bar \beta }_{{\xi _K}}})}}{h^{new}} \approx 0.
\end{align}
So far,  we have verified that the new transformation  $\delta_{(\xi,\beta_\xi,\bar\beta_\xi)}$ proposed in previous subsection indeed can remedy the non-zero Lie derivative of new fields on field space with non-zero $\varepsilon$. Note that when we choose the solution $\bar \beta_\xi=0$, the new transformation ${\delta _{(\xi ,{\beta _\xi },{{\bar \beta }_\xi })}}$ reduces to the Kosmann derivative. These results imply that, unlike for the Kosmann derivative, the field space transition from $(\tilde \omega,\tilde e)$ to $(\hat\omega,\tilde e,h)$  for ensuring the closed algebras is determined by different solutions of $\bar \beta_\xi$ in the first order equation \eqref{db}. Thus, we claim  that the Lorentz vector $\bar \beta_\xi$ quantifies the degree of incompleteness of the translation symmetry. This statement is based on the following facts. By rescaling $\lambda_\xi$ and neglecting the terms at order ${\cal O}((1+\sigma)^2)$, the closed algebra \eqref{operator algebra two} fails to be valid. In this sense, the translation symmetry becomes incomplete on the field space $(\tilde \omega,\tilde e)$, and therefore a field space that simultaneously supports transformations   $\delta_\varphi^\text{t}$, $\delta_\varphi^\text{m}$, and the closed algebras cannot exist, as discussed in Section \ref{section two}.

Consequently, the total charge $D_\xi^N$  including  an auxiliary charge $Q_{\bar \beta_\xi}^{\text m}$ is given by  the modified description
\begin{align}\label{DN charge btz}
    D_\xi^N=D_\xi(e,\omega,h)+Q_{  {{ \beta }_\xi }}^{\rm{j}} + Q_{{{\bar \beta }_\xi }}^{\rm{m}}.
\end{align}
which is just \eqref{new charge}. According to  \eqref{gauge charges}, the terms in  \eqref{DN charge btz}  are
\begin{align}\label{total new charge}
Q_{{\beta _\xi }}^{\rm{j}} + Q_{{{\bar \beta }_\xi }}^{\rm{m}} = \int\limits_{\cal S} {{\beta _\xi }(\frac{1}{\mu }\omega  - \sigma e) + {{\bar \beta }_\xi }e}= \int\limits_{\cal S} {(1 + \sigma \alpha )\left( {\varepsilon_1 (\sigma e^1-\frac{1}{\mu }\omega^1  ) + {{\bar \beta }_\xi }e} \right) - \frac{\alpha }{\mu }{{\bar \beta }_\xi }\omega } .
\end{align}

Specially, in the case with $\varepsilon=0$, i.e., $\lambda_\xi=0$, we have $\beta_\xi=-\alpha\bar\beta_\xi$, then the charges above are
\begin{align}
    Q_{{\beta _\xi }}^{\rm{j}} + Q_{{{\bar \beta }_\xi }}^{\rm{m}} = 0\quad \mathrm{and}\quad D_\xi^N=D_\xi
\end{align}
where we have used \eqref{boundary condition of one-form}. Thus, imposing condition \eqref{boundary condition of one-form} on the Lorentz vector not only ensures that $D_\xi^N$ properly reduces to $D_\xi$, but also makes the $\bar \beta_\xi$ in \eqref{path order solution} depend solely on the integral of the field-independent quantity $\varepsilon$. This means that $\bar \beta_\xi$ is also field-independent, which in turn renders the two gauge charges in \eqref{total new charge} integrable.

In the limit $\mu\to\infty$, the charge $D_\xi^N$ becomes approximately gauge-invariant when the integral proportional to $1+\sigma\alpha$ are ignored. This is a remarkable characteristic emerging from  the new  transformation  \eqref{new variation}.

\end{itemize}

\section{Conclusions and discussions}
The three-dimensional MMG introduces an auxiliary field, $h$. The parameters in this theory split the solutions into two families based on whether $1+\sigma\alpha=0$. For the case where this condition holds, a specific solution for the field $h$ is provided. We focused on the gauge symmetries, exact symmetries and the corresponding conserved charges in MMG, the study of which could provide insights into holographic renormalization, asymptotic symmetries in three-dimensional gravity theories that incorporate auxiliary fields.

We found that MMG possesses three independent gauge symmetries: Lorentz transformation, spacetime translation, and a new translation related to the field $h$. To ensure these symmetries form a algebraic structure, the field spaces were redefined, and we constructed two distinct closed algebras \eqref{operator algebra two} and \eqref{m algebras} in this framework. In the first, Lorentz transformation and standard translation act on the redefined fields $(\tilde \omega, \tilde e)$. In the second, Lorentz transformation and the new translation act on an expanded set of fields $(\hat \omega, \tilde e, h)$. This revealed a novel property of MMG that its complete symmetry algebra is not fixed but depends on the choice of parameters. It is worthwhile to note that our studies  solve the problem of the algebra not being closed in the original solution space.

We constructed the gauge charges corresponding to the three types of gauge symmetries with the use of covariant phase space method, and a new auxiliary charge was introduced comparing against previous theories. We verified that the three gauge charges and the diffeomorphism charge associated with the Lie derivative are equivalent under the on-shell case.

Further, by incorporating the mixed transformations involving $\delta_\varphi^\text{j}$ and $\delta_\varphi^\text{m}$, we found that the Kosmann derivative should be extended to be a new transformation ${\delta _{(\xi ,{\beta _\xi },{{\bar \beta }_\xi })}}$ (see \eqref{new variation}) in the sense of generating exact symmetries to allow approximate closure of algebras for small $1+\sigma\alpha$. This newly emerged transformation is unique to $\rm{MMG}$, which is featured by the auxiliary field $h$. However, defining such transformations is inherently challenging because determining the Lorentz vectors $\beta_\xi$ and $\bar \beta_\xi$ requires that the new transformation performing on all three types of fields must vanish identically when the chosen vector is a Killing vector. To achieve this, we proposed the conditions  \eqref{condition on new variation} which  are also compatible with the equations of motion or integrable conditions,  giving an extra constraint reflected in the parametric space. To better illustrate the distinctions between this new transformation and the usual transformations, we provided a detailed procedure using the rotational BTZ black hole as an example, in which we solved $\beta_\xi$ and $\bar \beta_\xi$ based on \eqref{db} by a path-ordered exponential in small $(1+\sigma\alpha)$ limit. Based on different solutions of the Lorentz vector $\bar \beta_\xi$, we found that only approximately closed algebras are allowed on extended field space $(\hat\omega,\tilde e,h)$. This means the constrained parameters and different solutions of $\bar \beta_\xi$ measure the approximate completeness of the translation symmetry. Finally, a new charge \eqref{new charge} associated with the new transformation arose. Our work introduced a novel mechanism to generate exact symmetries in theories with auxiliary fields, offering an approach that goes beyond conventional methods.

Based on our findings, we would suggest the following  future potential directions:
\begin{itemize}
\item In other three-dimensional gravity theories formulated within the triad formalism, such as those considered in \cite{Merbis:2014vja,Bergshoeff:2009zz,deRham:2010ik,deRham:2010kj,Sinha:2010ai}, the space field redefinition techniques employed in this work may likewise be implemented to construct additional closed algebraic structures. Recently, there had been several relevant studies concerning extended symmetries \cite{Gallagher:2023ghl, Andrianopoli:2023dfm,Borsten:2024pfz}. Furthermore, the method developed herein is expected to be applicable to a broader class of non-topological black hole solutions in MMG, including those presented in \cite{Alishahiha:2014dma}, where the auxiliary field $h$ is no longer constrained to be proportional to the triad. By adopting analogous procedures, one can systematically investigate the corresponding thermodynamic properties and integrability conditions of such solutions.
\item The emergence of novel symmetry in our analysis, which preserves the metric, motivates a generalization of the traditionally considered Lorentz-invariant fields in the Chern-Simons representation of three-dimensional gravity to fields that are compatible with the novel symmetry. This generalization facilitates the derivation of the most comprehensive set of boundary conditions for the three-dimensional case, as discussed in \cite{Grumiller:2016pqb}, especially for leaky boundary conditions  \cite{Fiorucci:2021pha, McNees:2024iyu} along with a complete classification of the associated asymptotic symmetries. A similar line of generalization may also be pursued in the two-dimensional context of \cite{Grumiller:2017qao}. However, this requires a detailed investigation into the origin of the new symmetries from auxiliary fields and the structure of the resulting Casimir operators \cite{Gonzalez:2018enk} in lower dimensions. We are also curious about how our new symmetry is affected by corner constraints \cite{Freidel:2020svx}, as the new degrees of freedom should undergo some form of reduction to give rise to the new symmetry \cite{Freidel:2020ayo}.
\item Since the Kosmann derivative ${\cal K}_\xi$ may violate the triad gauge condition, it necessitates additional compensating transformations to restore gauge compatibility. For instance, consider a gauge choice that fixes the component ${e_x}^1$ on the boundary. In general, the operation of ${\cal K}_\xi$  does not preserve this condition, i.e., ${{ {\cal K}}_\xi } {e_x}^1\neq 0$. To compensate for this deviation, an additional transformation generated by a gauge-dependent  $\bar \lambda_\xi$ must be introduced such that ${{\cal K}_\xi }{e_x}^1 + \delta _{\bar \lambda_\xi}^{\text{j}}{e_x}^1 = 0$. This leads to a modified expression for the Kosmann derivative that is compatible with the chosen gauge of the triad and ambiguities of corners \cite{Geiller:2021vpg}. Having confirmed in our analysis that the newly introduced symmetry operators satisfy  the first law of black hole thermodynamics, we expect to adapt the one-forms associated with coordinate gauge freedom to incorporate the new transformations developed in this work. This extension is expected to facilitate applications to more general solutions of the Lorentz vector $\bar\beta_\xi$, including non-stationary solutions as considered in \cite{Geiller:2021vpg,Ciambelli:2024vhy,Poole:2018koa,Compere:2023ktn}.

\end{itemize}

\section*{Acknowledgment}
We extend our sincere gratitude to Prof. Weijia Li and Lei Li from Dalian University of Technology for their indispensable discussions. This work is partly supported by the Natural Science Foundation of China under Grants No. 12375054 and the Postgraduate Research $\&$ Practice Innovation Program of Jiangsu Province under Grants No. KYCX-3502.

\appendix
\section{Notation and convention}
\label{appendix one}
In this appendix, we shall clarify some notations and conventions to better explain the Lagrangian \eqref{main L} and rules of calculations. We choose $\mu,\nu,\alpha\in \left\{ {0,1,2,\cdots} \right\}$ and   $I,J,K,\cdots \in \left\{ {0,1,2,\cdots} \right\}$ as coordinate index and abstract index, respectively, and  $ \eta^{IJ}$ as the internal metric in form $\label{internal metric}
      {\eta ^{IJ}} = \left( {\begin{array}{*{20}{c}}
-1&0&0\\
0&1&0\\
0&0&1
\end{array}} \right)$. In our notation, we define anti-commutation of coordinate index ${A_{\left[ \mu  \right.}}{B_{\left. \nu  \right]}} = {A_\mu }{B_\nu } - {A_\nu }{B_\mu }$ and introduce two products $\cdot\wedge\cdot$ and $[\cdot \wedge \cdot]$
\begin{align}
	\begin{aligned}
&{P_1} \wedge {P_2} = \frac{{({p_1} + {p_2})!}}{{{p_1}!{p_2}!}}{\eta _{IJ}}{P_1}_{[{\mu _1} \cdots {\mu _{{p_1}}}}^I{P_2}_{{\mu _{{p_1} + 1}} \cdots {\mu _{{p_1} + {p_2}]}}}^J{\tilde\varepsilon ^{{\mu _1} \cdots {\mu _{{p_1} + {p_2}}}}}{{\rm{d}}^{{p_1} + {p_2}}}x,\\
&\left[ {{P_1} \wedge {P_2}} \right] = {\varepsilon ^I}_{JK}{P_1}^J{P_2}^K = \frac{{({p_1} + {p_2})!}}{{{p_1}!{p_2}!}}{\varepsilon ^I}_{JK}{P_1}_{[{\mu _1} \cdots {\mu _{{p_1}}}}^J{P_2}_{{\mu _{{p_1} + 1}} \cdots {\mu _{{p_1} + {p_2}]}}}^K{\tilde\varepsilon ^{{\mu _1} \cdots {\mu _{{p_1} + {p_2} + 1}}}}{{\rm{d}}^{{p_1} + {p_2}}}x,
	\end{aligned}
\end{align}
 where $P_i$ is a $p_i$-form and written as
 \begin{align}
 P_i=\frac{1}{p_{i}!} P_{i \mu_1 \cdots \mu_{p i}} \mathrm{~d} x^{\mu_1} \wedge \cdots \wedge \mathrm{~d} x^{\mu_{p i}}.
 \end{align}
 The explicit expression of some three-forms in the Lagrangian are
\begin{align}
	\begin{aligned}
			&  \sigma e \wedge R(\omega) =   {\sigma} {e_\mu} ^I{({\text d}{\omega) _{\nu \alpha }}_I}{\tilde\varepsilon ^{\mu \nu \alpha }}{\text d^3 x} +{\sigma }{\tilde\varepsilon ^I}_{JK}{e_{\mu I }}{\omega _\nu} ^{J}{\omega _\alpha }^K{\tilde\varepsilon ^{\mu \nu \alpha }}{\text d^3 x},\\
		&	e \wedge \left[ {e \wedge e} \right] =2 {\varepsilon ^I}_{JK}{e_{\mu I }}{e_\nu} ^{J}{e_\alpha} ^K{\tilde\varepsilon ^{\mu \nu \alpha }}{\text d^3 x},
	\end{aligned}
\end{align}
where the Levi-Civita symbol $\varepsilon_{IJK}$ with abstract index is risen by the internal metric $\eta^{IJ}$ and follows $\varepsilon_{012}=1$ and $\text{d}^3x={\rm{d}}{x^1} \wedge {\rm{d}}{x^2} \wedge {\rm{d}}{x^3}$. The real tensor $\varepsilon_{\mu\nu\alpha}$ is defined as $\varepsilon_{\mu\nu\alpha}=\sqrt{\left| g \right|}\tilde\varepsilon_{\mu\nu\alpha}$, where $\tilde\varepsilon_{012}=1$, and its indexes are risen by the metric. Two useful contractions between $\varepsilon_{IJK}$ and $\varepsilon^{IJK}$ are displayed here
\begin{subequations}
    \begin{align}
&{\varepsilon _{IJK}}{\varepsilon ^{ILM}} = \delta _K^L\delta _J^M - \delta _J^L\delta _K^M,\\
&{\varepsilon _{IJK}}{\varepsilon ^{IJL}} =  - 2\delta _K^L.
    \end{align}
\end{subequations}

The commutation relationship of $P_1$ and $P_2$ in these two brackets are
\begin{align}\label{A.3}
	P_1 \wedge P_2 = {( - 1)^{p_1p_2}}P_2 \wedge P_1,\quad \quad \left[ {P_1 \wedge P_2} \right] = {( - 1)^{p_1p_2 + 1}}\left[ {P_2 \wedge P_1} \right].
\end{align}
There is also a relationship for the mixed bracket
\begin{align}
	\left[ {P_1 \wedge P_2} \right] \wedge P_3 = {( - 1)^{(p_1 + p_2)p_3}}\left[ {P_3 \wedge P_1} \right] \wedge P_2.
\end{align}
We also introduce two types derivative
\begin{align}
\text{d}_\omega P_1=\text{d} P_1+\left[ {\omega  \wedge {P_1}} \right],\quad  \text{d}_\Omega P_1=\text{d} P_1+\left[ {\Omega  \wedge {P_1}}\right].
\end{align}
For $\text{d}_\omega$ and $\text{d}$, the Leibniz rule for products in \eqref{A.3} follow
\begin{subequations}
\begin{align}
	&{\text{d}_\omega }\left[ {P_1 \wedge P_2} \right] = \left[ {{\text{d}_\omega }P_1 \wedge P_2} \right] + {( - 1)^{p_1}}\left[ {P_1 \wedge {\text{d}_\omega }P_2} \right],\\
    &\text{d}(P_1 \wedge P_2) = {\text{d}_\omega }P_1 \wedge P_2 + {( - 1)^{p_1 }}P_1 \wedge {\text{d}_\omega }P_2.
\end{align}
\end{subequations}

Using the equations of motion \eqref{EOM}, ${\text d}_{\omega}i_\xi e,{\text d}_{\omega}i_\xi \omega$ and ${\text d}_{\omega}i_\xi h$ can be expressed as
\begin{subequations}\label{domega on fields}
    \begin{align}
&{{\rm{d}}_\omega }({i_\xi }\omega ) = {{\cal L}_\xi }\omega  - {i_\xi }R(\omega ) \approx {{\cal L}_\xi }\omega  + \mu (1 + \sigma \alpha )\left[ {{i_\xi }e,h} \right] - \mu (1 + \sigma \alpha )\left[ {e,{i_\xi }h} \right],\\
&{{\rm{d}}_\omega }({i_\xi }e) = {{\cal L}}_\xi e - {i_\xi }T(e) + \left[ {{i_\xi }\omega ,e} \right] \approx {L_\xi }e + \alpha \left[ {{i_\xi }e,h} \right] - \alpha \left[ {e,{i_\xi }h} \right] + \left[ {{i_\xi }\omega ,e} \right],\\
&{{\rm{d}}_\omega }({i_\xi }h) = {{\cal L}_\xi }h - {i_\xi }{{\rm{d}}_\omega }h + \left[ {{i_\xi }\omega ,h} \right] \approx {L_\xi }h - \alpha \left[ {{i_\xi }h,h} \right] - {\Lambda _0}\left[ {{i_\xi }e,e} \right] + \left[ {{i_\xi }\omega ,h} \right]\notag\\
 &- \sigma \mu (1 + \sigma \alpha )\left[ {{i_\xi }e,h} \right] + \sigma \mu (1 + \sigma \alpha )\left[ {e,{i_\xi }h} \right].
    \end{align}
\end{subequations}

\section{Some explicit calculations of deriving gauge symmetries and operator algebras}
\label{appendix 3}

\subsection{Derivation of \eqref{action on L}}\label{sec:B-1}
We  elucidate the derivation of Equation \eqref{action on L} in this part. We begin with a general transformation $\delta _\varphi ^{tol}$ consisting of $\delta _\varphi ^{\text{j}}$, $\delta _\varphi ^{\text{t}}$ and $\delta _\varphi ^{\text{m}}$ with constants $a$, $b$ and $c$
\begin{align}\label{b.1}
	\delta _\varphi ^{tol} = a\delta _\varphi ^{\text{j}} + b\delta _\varphi ^{\text{t}} + c\delta _\varphi ^{\text{m}},
\end{align}
which operates on Lagrangian $L$, (see \eqref{L} and \eqref{EOM}), yielding
\begin{align}
\begin{aligned}\label{variation L}
&\delta _\varphi ^{total}L - d(aI_\varphi ^{\text{j}}\theta  + bI_\varphi ^{\text{t}}\theta  + cI_\varphi ^{\text{m}}\theta ) = {E_h} \wedge \left\{ {a[h,\varphi ] + b(\sigma \mu (1 + \sigma \alpha )[h,\varphi ] + {\Lambda _0}[e,\varphi ])} \right.\\
&\left. { + c({{\rm{d}}_\omega }\varphi  + \alpha [h,\varphi ] + \sigma \mu (1 + \sigma \alpha )[e,\varphi ])} \right\} + \frac{{E_\omega }}{\mu} \wedge \left\{ {a{{\rm{d}}_\omega }\varphi  + b\mu (1 + \sigma \alpha )[h,\varphi ]} \right.\\
&\left. { + c\mu (1 + \sigma \alpha )[e,\varphi ]} \right\} + {E_e} \wedge \left\{ {(a + c\alpha )[e,\varphi ] + b{{\rm{d}}_\omega }\varphi  + b\alpha [h,\rho ]} \right\},\\
 &= T(e) \wedge \left\{ {(a + b\sigma \mu (1 + \sigma \alpha ) + c\alpha )[h,\varphi ]} \right.\left. { + (b{\Lambda _0} + c\sigma \mu (1 + \sigma \alpha ))[e,\varphi ] + c{{\rm{d}}_\omega }\varphi } \right\}\\
& + c\alpha [e \wedge h] \wedge {{\rm{d}}_\omega }\varphi  + R(\omega ) \wedge \left\{ {\frac{a}{\mu}{{\rm{d}}_\omega }\varphi  + b (1 + \sigma \alpha )[h,\varphi ] + c (1 + \sigma \alpha )[e,\varphi ]} \right\}\\
& + a [e \wedge h] \wedge {{\rm{d}}_\omega }\varphi  - \sigma  T(e) \wedge \left\{ {a{{\rm{d}}_\omega }\varphi  + b\mu (1 + \sigma \alpha )[h,\varphi ] + c\mu (1 + \sigma \alpha )[e,\varphi ]} \right\}\\
& + \frac{\alpha }{2}b[h \wedge h] \wedge {{\rm{d}}_\omega }\varphi  - \sigma R(\omega ) \wedge \left\{ {(a + c\alpha )[e,\varphi ] + b{{\rm{d}}_\omega }\varphi  + b\alpha [h,\rho ]} \right\}\\
& + \frac{{\sigma {\Lambda _0}}}{2}[e \wedge e] \wedge {{\rm{d}}_\omega }\varphi  + {{\rm{d}}_\omega }h \wedge \left\{ {(a + c\alpha )[e,\varphi ] + b{{\rm{d}}_\omega }\varphi  + b\alpha [h,\rho ]} \right\}.
 \end{aligned}
\end{align}
In the second equality above,  we assumed the contraction ${e_\mu }^I{h_\nu }_I$  to be a symmetric tensor $h_{\mu\nu}$\footnote{Note that the validity of this condition inherently relies on the specific form of $h$ obtained from the equations of motion. For generality, only solutions satisfying this condition are analyzed here.}. Now, to ensure $\delta _\varphi ^{total}L$ becomes a total derivative term, we select parameters $a$, $b$ and $c$.
Further considering that
\begin{subequations}
    \begin{align}
&R(\omega ) \wedge {\text{d}_\omega }\varphi  = \text{d}(R(\omega ) \wedge \varphi ),\\
      &T(e) \wedge {\text{d}_\omega }\varphi  = \text{d}(T(e) \wedge\varphi ) + \left[ {\varphi  \wedge e} \right] \wedge R(\omega),\\
        &{\text{d}_\omega }h \wedge {\text{d}_\omega }\varphi  = \text{d}(h \wedge {\text{d}_\omega }\varphi ) + h \wedge {\text{d}_\omega }{\text{d}_\omega }\varphi  = (h \wedge {\text{d}_\omega }\varphi ) + h \wedge \left[ {R(\omega ) \wedge \varphi } \right],\\
        &\left[ {e \wedge h} \right] \wedge {\text{d}_\omega }\varphi  = \text{d}(\left[ {e \wedge h} \right] \wedge \varphi ) + \left[ {\varphi  \wedge h} \right] \wedge {\text{d}_\omega }e + \left[ {\varphi  \wedge e} \right] \wedge {\text{d}_\omega }h,
    \end{align}
\end{subequations}
help to simplify  \eqref{variation L} to
\begin{align}\label{eq:total delta L}
	\begin{aligned}
		&\delta _\varphi ^{total}L-d(aI_\varphi ^{\text{j}}\theta  + bI_\varphi ^{\text{t}}\theta  + cI_\varphi ^{\text{m}}\theta ) = T(e) \wedge \left\{ {\left( { b\sigma \mu (1 + \sigma \alpha )  - \sigma b{\mu }(1 + \sigma \alpha )} \right)[h,\varphi ]} \right.\\
		&\left. { + \left( {c\sigma \mu (1 + \sigma \alpha ) - \sigma c{\mu }(1 + \sigma \alpha )} \right)[e,\varphi ]} \right\} + R(\omega ) \wedge \left\{ {\left( {b (1 + \sigma \alpha ) - \sigma b\alpha  - b} \right)[h,\varphi ]} \right.\\
		&\left. { + \left( {c\mu (1 + \sigma \alpha ) - \sigma (a + c\alpha ) + \sigma \mu a - c} \right)[e,\varphi ]} \right\} + {\text{d}_\omega }h \wedge \left\{ {a(1 - \mu )[e,\varphi ]} \right\}+ \text{d}B_\varphi
	\end{aligned}
\end{align}
with
\begin{align}\label{Bf}
	\begin{aligned}
		&{B_\varphi } = (\frac{a}{\mu} - \sigma b)R(\omega ) \wedge \varphi  + (c\alpha  + a )\varphi  \wedge \left[ {e \wedge h} \right] + (c - \sigma  a)T(e) \wedge \varphi  + \frac{\alpha }{2}b\varphi  \wedge \left[ {h \wedge h} \right]\\
		& + \frac{{b{\Lambda _0}}}{2}\varphi  \wedge \left[ {e \wedge e} \right] + bh \wedge {\text{d}_\omega }\varphi.
	\end{aligned}
\end{align}
It is direct to get that
\begin{align}\label{actions on L}
	\delta _\varphi ^{total}L-\text{d}(aI_\varphi ^{\text{j}}\theta  + bI_\varphi ^{\text{t}}\theta  + cI_\varphi ^{\text{m}}\theta ) = \text{d}{B_\varphi},
\end{align}
implying that  $\delta _\varphi ^{total}$ generates gauge symmetries and one does not use equations of motion. Moreover,  the cases with  $a=1$ and $ b=c=0$; $b=1$ and $a=c=0$; and $c=1$ and $a=b=0$  yield the three sub-equations in \eqref{action on L}, respectively.

\subsection{Discussions on  \eqref{operator algebra two}}\label{sec:B-2}
In this part, we clarify the properties of  the closed algebras  \eqref{operator algebra two}  formed by the transformations in the new space fields \eqref{larger space}, and show the failure to incorporate the new transformation $\delta_\varphi^{\text{m}}$ into a closed algebra.
The operation of $\delta_\chi^{\text{m}}$  on $\tilde{\omega}$ and $\tilde{e}$ gives
\begin{subequations}
	\begin{align}
		& \delta_\chi^{\text{m}} \tilde{\omega}=\alpha a_1[e, \chi]+\mu(1+\sigma \alpha)[h, \chi]+c_1\left(\text{d}_\omega \chi+\alpha[h, \chi]+\sigma \mu(1+\sigma \alpha)[e, \chi]\right), \\
		& \delta_\chi^{\text{m}} \tilde{e}=\alpha a_2[e, \chi]+c_2\left(\text{d}_\omega \chi+\alpha[h, \chi]+\sigma \mu(1+\sigma \alpha)[e, \chi]\right),
	\end{align}
\end{subequations}
respectively. To ensure closure, we first consider that \eqref{eqalgebra:total} with the parameters \eqref{b.8constraints} is satisfied via field-independent zero-forms $\varphi_1$ and $\varphi_2$. Subsequently, the commutator $\left[ \delta_\tau^{\text{j}}, \delta_\chi^{\text{m}} \right]$ on $\tilde{\omega}$ is
\begin{subequations}
	\begin{align}\label{jm omega}
		\delta _{{\varphi _2}}^{\text{j}}\delta _{{\varphi _1}}^{\text{m}}\tilde \omega&  = {a_1}\alpha \left[ {\left[ {e,{\varphi _2}} \right],{\varphi _1}} \right] + \mu (1 + \sigma \alpha )\left[ {\left[ {h,{\varphi _2}} \right],{\varphi _1}} \right] \notag \\&+ {c_1}\left( {\left[ {{\text{d}_\omega }{\varphi _2},{\varphi _1}} \right] + \alpha \left[ {\left[ {h,{\varphi _2}} \right],{\varphi _1}} \right] + \sigma \mu (1 + \sigma \alpha )\left[ {\left[ {e,{\varphi _2}} \right],{\varphi _1}} \right]} \right),\\
		\delta _{{\varphi _1}}^{\text{m}}\delta _{{\varphi _2}}^{\text{j}}\tilde \omega & = {a _1}\alpha \left[ {\left[ {e,{\varphi _1}} \right],{\varphi _2}} \right] + \mu (1 + \sigma \alpha )\left[ {\left[ {h,{\varphi _1}} \right],{\varphi _2}} \right] \notag \\&+ {c_1}\left( {\left[ {{\text{d}_\omega }{\varphi _1},{\varphi _2}} \right] + \alpha \left[ {\left[ {h,{\varphi _1}} \right],{\varphi _2}} \right] + \sigma \mu (1 + \sigma \alpha )\left[ {\left[ {e,{\varphi _1}} \right],{\varphi _2}} \right]} \right),
	\end{align}
\end{subequations}
and parallel procedures for $\tilde{e}$ and $h$ yields
\begin{subequations}
	\begin{align}
		\delta _{{\varphi _2}}^{\text{j}}\delta _{{\varphi _1}}^{\text{m}}\tilde e &=\alpha {a_2}\left[ {\left[ {e,{\varphi _2}} \right],{\varphi _1}} \right]+ {c_2}\left[ {{\text{d}_\omega }{\varphi _2} + \alpha \left[ {h,{\varphi _2}} \right] + \sigma \mu (1 + \sigma \alpha )\left[ {e,{\varphi _2}} \right],{\varphi _1}} \right],\\
		\delta _{{\varphi _1}}^{\text{m}}\delta _{{\varphi _2}}^{\text{j}}\tilde e& =\alpha {a_2}\left[ {\left[ {e,{\varphi _1}} \right],{\varphi _2}} \right]+ {c_2}\left[ {{\text{d}_\omega }{\varphi _1} + \alpha \left[ {h,{\varphi _1}} \right] + \sigma \mu (1 + \sigma \alpha )\left[ {e,{\varphi _1}} \right],{\varphi _2}} \right],\\
		\delta _{{\varphi _1}}^{\text{j}}\delta _{{\varphi _2}}^{\text{m}}h& = \left[ {{\text{d}_\omega }{\varphi _1},{\varphi _2}} \right] + \alpha \left[ {\left[ {h,{\varphi _1}} \right],{\varphi _2}} \right] + \sigma \mu (1 + \alpha \sigma )\left[ {\left[ {e,{\varphi _1}} \right],{\varphi _2}} \right]\label{jm h1},\\
		\delta _{{\varphi _2}}^{\text{m}}\delta _{{\varphi _1}}^{\text{j}}h& = \left[ {{\text{d}_\omega }{\varphi _2},{\varphi _1}} \right] + \alpha \left[ {\left[ {h,{\varphi _2}} \right],{\varphi _1}} \right]+ \sigma \mu (1 + \alpha \sigma )\left[ {\left[ {e,{\varphi _2}} \right],{\varphi _1}} \right]\label{jm h2},
	\end{align}
\end{subequations}
which implies that  there is a closed algebra $\left[ \delta_\tau^{\text{j}}, \delta_\chi^{\text{m}} \right] = \delta_{[\tau, \chi]}^{\text{m}}$ on $(\tilde \omega,\tilde e,h)$.

Then we check the larger algebra including $\delta _\varphi ^{\text{t}}$. The operations   $\delta _{\varphi_1} ^{\text{m}}\delta _{\varphi_2} ^{\text{t}}$ and $\delta _{\varphi_2} ^{\text{t}}\delta _{\varphi_1} ^{\text{m}}$ on $\tilde \omega$ are
\begin{subequations}
	\begin{align}\label{tm omega}
		\delta _{{\varphi _2}}^{\text{t}}\delta _{{\varphi _1}}^{\text{m}}\tilde \omega&  = (\mu (1 + \sigma \alpha ) + {c_1}\alpha )\left[ {\sigma \mu (1 + \sigma \alpha )\left[ {h,{\varphi _2}} \right] + {\Lambda _0}\left[ {e,{\varphi _2}} \right],{\varphi _1}} \right]\notag\\& + {c_1}\mu (1 + \sigma \alpha )\left[ {\left[ {h,{\varphi _2}} \right],{\varphi _1}} \right] + {c_1}\sigma \mu (1 + \sigma \alpha )\left[ {{\text{d}_\omega }{\varphi _2} + \alpha \left[ {h,{\varphi _2}} \right],{\varphi _1}} \right],\\
		\delta _{{\varphi _1}}^{\text{m}}\delta _{{\varphi _2}}^{\text{t}}\tilde \omega & = {a_3}\left[ {\alpha {a_2}\left[ {e,{\varphi _1}} \right] + {c_2}({\text{d}_\omega }{\varphi _1} + \alpha [h,{\varphi _1}] + \sigma \mu (1 + \sigma \alpha )[e,{\varphi _1}]),{\varphi _2}} \right],
	\end{align}
\end{subequations}
and on $\tilde e$ give
\begin{subequations}
	\begin{align}\label{tm e}
		\delta _{{\varphi _2}}^{\text{t}}\delta _{{\varphi _1}}^{\text{m}}\tilde e &= \left( {{a_2}\alpha  + {c_2}\sigma \mu (1 + \sigma \alpha )} \right)\left[ {{\text{d}_\omega }{\varphi _2} + \alpha [h,{\varphi _2}],{\varphi _1}} \right] + {c_2}\alpha \mu (1 + \sigma \alpha )\left[ {[h,{\varphi _2}],{\varphi _1}} \right]\notag\\
		& + {c_2}\alpha \sigma \mu (1 + \sigma \alpha )\left[ {[h,{\varphi _2}],{\varphi _1}} \right]+ {c_2}\alpha {\Lambda _0}\left[ {[e,{\varphi _2}],{\varphi _1}} \right],\\
		\delta _{{\varphi _1}}^{\text{m}}\delta _{{\varphi _2}}^{\text{t}}\tilde e& = {a_2}\left[ {\mu (1 + \sigma \alpha )[h,{\varphi _1}] + {c_1}\left( {{\text{d}_\omega }{\varphi _1} + \alpha [h,{\varphi _1}] + \sigma \mu (1 + \sigma \alpha )[e,{\varphi _1}]} \right),{\varphi _2}} \right]\notag\\
		& + {b_3}[\alpha {a_2}[e,{\varphi _1}]+ {c_2}\left( {{\text{d}_\omega }{\varphi _1} + \alpha [h,{\varphi _1}] + \sigma \mu (1 + \sigma \alpha )[e,{\varphi _1}]} \right),{\varphi _2}].
	\end{align}
\end{subequations}
According to \eqref{tm omega} and  \eqref{tm e}, the operations yield closure of algebras only when the parameters fulfill the relations
\begin{align}\label{b.16}
	\begin{aligned}
&(\mu (1 + \sigma \alpha ) + 2{c_1}\alpha )\sigma \mu (1 + \sigma \alpha ) + {c_1}\mu (1 + \sigma \alpha ) = {a_3}{c_2}\alpha, \\
&(\mu (1 + \sigma \alpha ) + {c_1}\alpha ){\Lambda _0} = {a_3}\left( {{a_2}\alpha  + {c_2}\sigma \mu (1 + \sigma \alpha )} \right),\\
&{c_1}\sigma \mu (1 + \sigma \alpha ) = {a_3}{c_2},\\
&{a_2}\alpha  + {c_2}\sigma \mu (1 + \sigma \alpha ) = {a_2}{c_1} + {b_3}{c_2},\\
&{c_2}^2{\alpha ^2}\sigma \mu (1 + \sigma \alpha ) = {a_2}{c_1}\sigma \mu (1 + \sigma \alpha ) + {b_3}\alpha {a_2} + {b_3}{c_2}\sigma \mu (1 + \sigma \alpha ),\\
&\alpha \left( {{a_2}\alpha  + {c_2}\sigma \mu (1 + \sigma \alpha )} \right) + {c_2}\alpha \mu (1 + \sigma \alpha ) + {c_2}\alpha \sigma \mu (1 + \sigma \alpha ) = {a_2}\mu (1 + \sigma \alpha ) + {c_1}\alpha  + {b_3}{c_2}\alpha,
	\end{aligned}
\end{align}
which generally contradict with the classification below \eqref{eqalgebra:total}. These constraints are not acceptable because the parameters are over-determined. Therefore,  we argue that the transformation $\delta _{{\varphi}}^{\text{m}}$ cannot form a closed algebra on space $(\tilde \omega, \tilde e)$ when $\delta_m^{\text{t}}$ is involved.

\subsection{Derivation of \eqref{m algebras} }\label{sec:B-3}
The double operations of $\delta^{\text{m}}_\varphi$ on $\tilde{\omega}$, $\tilde{e}$, and $h$ are
\begin{subequations}
	\begin{align}
		\delta _{{\varphi _1}}^{\text{m}}\delta _{{\varphi _2}}^{\text{m}}\tilde \omega & = \left( {{c_1}\alpha \sigma \mu (1 + \sigma \alpha ) + \sigma {\mu ^2}{{(1 + \sigma \alpha )}^2} + {c_1}\sigma \alpha \mu (1 + \sigma \alpha )} \right)\left[ {\left[ {e,{\varphi _1}} \right],{\varphi _2}} \right]\notag\\
		&+ \left( {{c_1}{\alpha ^2} + \mu \alpha (1 + \sigma \alpha ) + {c_1}\mu (1 + \sigma \alpha )} \right)\left[ {\left[ {h,{\varphi _1}} \right],{\varphi _2}} \right]\notag\\& + \left( {{c_1}\alpha  + \mu (1 + \sigma \alpha )} \right)\left[ {{\text{d}_\omega }{\varphi _1},{\varphi _2}} \right],\\
		\delta _{{\varphi _1}}^{\text{m}}\delta _{{\varphi _2}}^{\text{m}}\tilde e& = \left( {{a_2}{\alpha ^2} + \sigma \mu \alpha (1 + \sigma \alpha ) + {c_2}\sigma \mu \alpha (1 + \sigma \alpha )} \right)\left[ {\left[ {e,{\varphi _1}} \right],{\varphi _2}} \right] + {c_2}\alpha \left[ {{\text{d}_\omega }{\varphi _1},{\varphi _2}} \right]\notag\\
		&+ \left( {{c_2}\mu (1 + \sigma \alpha ) + {c_2}{\alpha ^2}} \right)\left[ {\left[ {h,{\varphi _1}} \right],{\varphi _2}} \right],\\
		\delta _{{\varphi _1}}^{\text{m}}\delta _{{\varphi _2}}^{\text{m}}h &= \alpha \left[ {{\text{d}_\omega }{\varphi _1},{\varphi _2}} \right] + (\mu (1 + \sigma \alpha ) + {\alpha ^2})\left[ {\left[ {h,{\varphi _1}} \right],{\varphi _2}} \right] + 2\alpha \sigma \mu (1 + \sigma \alpha )\left[ {\left[ {e,{\varphi _1}} \right],{\varphi _2}} \right]\label{mm h}.
	\end{align}
\end{subequations}
The result \eqref{mm h} gives the anti-commutation relation on $h$
\begin{align}
	\begin{aligned}
		&(\delta _{{\varphi _1}}^{\text{m}}\delta _{{\varphi _2}}^{\text{m}} - \delta _{{\varphi _2}}^{\text{m}}\delta _{{\varphi _1}}^{\text{m}})h = \mu (1 + \sigma \alpha )(1 - \frac{{\alpha \mu {\sigma ^2}(1 + \sigma \alpha )}}{{{\Lambda _0}}})\delta _{\left[ {{\varphi _1},{\varphi _2}} \right]}^{\text{j}}h  \\
		&+ \frac{{\mu\alpha \sigma (1 + \sigma \alpha )}}{{{\Lambda _0}}}\delta _{\left[ {{\varphi _1},{\varphi _2}} \right]}^{\text{t}}h+ \alpha \delta _{\left[ {{\varphi _1},{\varphi _2}} \right]}^{\text{m}}h,
	\end{aligned}
\end{align}
which cannot admit the closed algebra of $\delta _{{\varphi }}^{\text{m}}$ for general parameter. But for the parameters at point $1+\sigma\alpha=0$, it is direct to reduce
\begin{subequations}
	\begin{align}\label{closed algebra of h}
		&(\delta _{{\varphi _1}}^{\text{m}}\delta _{{\varphi _2}}^{\text{m}} - \delta _{{\varphi _2}}^{\text{m}}\delta _{{\varphi _1}}^{\text{m}})\tilde e = \alpha \delta _{\left[ {{\varphi _1},{\varphi _2}} \right]}^{\text{m}}\tilde e,\\
		&(\delta _{{\varphi _1}}^{\text{m}}\delta _{{\varphi _2}}^{\text{m}} - \delta _{{\varphi _2}}^{\text{m}}\delta _{{\varphi _1}}^{\text{m}})h = \alpha \delta _{\left[ {{\varphi _1},{\varphi _2}} \right]}^{\text{m}}h,\\
      &  (\delta _{{\varphi _1}}^{\text{m}}\delta _{{\varphi _2}}^{\text{m}} - \delta _{{\varphi _2}}^{\text{m}}\delta _{{\varphi _1}}^{\text{m}})\hat \omega  = \alpha \delta _{\left[ {{\varphi _1},{\varphi _2}} \right]}^{\text{m}}\hat \omega,
	\end{align}
\end{subequations}
with $\hat \omega=\frac{1}{\alpha}\tilde e+\omega$ , which suggest the closed algebra of $\delta _{{\varphi }}^{\text{m}}$ on space $(\hat\omega, \tilde e, h)$. Subsequently, combining with the derivations in Appendix \ref{sec:B-2}
 and \ref{sec:B-3} , we conclude that the transformations  $\delta_{\varphi}^{\text{j}}$ and $\delta_{\varphi}^{\text{m}}$ form closed algebras \eqref{m algebras} on the space $(\hat{\omega}, \tilde{e}, h)$.

\section{The proof of exact symmetries generated by $\delta_{(\xi ,\beta_{\xi},\bar\beta_{\xi})}$}
\label{appendix c}
In this appendix, the failure  to solve $\beta_{\xi_K}$ and $\bar\beta_{\xi_K}$ only through \eqref{51a} will be first elaborated. Then, we will verify  that the conditions  \eqref{condition on new variation} indeed can make $\delta_{({\xi_K} ,\beta_{\xi_K},\bar\beta_{\xi_K})}\omega\approx 0$ ,$\delta_{({\xi_K} ,\beta_{\xi_K},\bar\beta_{\xi_K})}h\approx 0$, and the constraint \eqref{parameters constraints} corresponds the integrable condition.

\subsection{Failure of expressing $\bar\beta_{\xi_K}$ in term of $\beta_{\xi_K}$ by \eqref{51a} and \eqref{51b1}}
\label{appendix c1}
For simplicity, we consider that both ${\cal L}_{\xi_K} e$, ${\cal L}_{\xi_K} \Omega$   vanish under on-shell condition and conditions \eqref{51a} and \eqref{51b1} hold. In this case, operating  $\delta_{(\xi_K ,\beta_{\xi_K},\bar\beta_{\xi_K})}$ on both sides of $E_h=0$ in \eqref{eom21} yields
\begin{align}\label{do}
    \left[ {(\delta _{{\beta _K}}^{\rm{j}} + \delta _{{{\bar \beta }_K}}^{\rm{m}})\Omega  \wedge e} \right] \approx 0,
\end{align}
 from which we have $(\delta _{{\beta _{{\xi _K}}}}^{\rm{j}} + \delta _{{{\bar \beta }_{{\xi _K}}}}^{\rm{m}})\Omega \approx0$  because the components of ${{e_\mu}^I}$ are linear independent.

To proceed, we use  \eqref{eq:Le-omega} and \eqref{gauge three} to rewrite $(\delta _{{\beta _{{\xi _K}}}}^{\rm{j}} + \delta _{{{\bar \beta }_{{\xi _k}}}}^{\rm{m}})\Omega $  as
\begin{align}\label{dom}
\begin{aligned}
&(\delta _{{\beta _{{\xi _K}}}}^{\rm{j}} + \delta _{{{\bar \beta }_{{\xi _K}}}}^{\rm{m}})\Omega  = {{\rm{d}}_\omega }{\beta _{{\xi _K}}} + \alpha \left[ {h,{{ \beta }_{{\xi _k}}}} \right] + \mu (1 + \sigma \alpha )\left[ {e,{\bar\beta _{{\xi _K}}}} \right]\\
& + \alpha \left( {{{\rm{d}}_\omega }{{\bar \beta }_{{\xi _K}}} + \alpha \left[ {h,{{\bar \beta }_{{\xi _K}}}} \right] + \sigma \mu (1 + \sigma \alpha )\left[ {e,{{\bar \beta }_{{\xi _K}}}} \right]} \right)\approx 0,
\end{aligned}
\end{align}
which becomes
\begin{align}
\begin{aligned}\label{debeta}
\left[ {e,{{\bar \beta }_{{\xi _K}}}} \right] \approx 0,
\end{aligned}
\end{align}
when the assumption ${\cal L}_{\xi_K} e\approx0$ or $\lambda_{\xi_K}\approx0$ is considered.

Considering that  $\delta_{({\xi _K},\beta_{{\xi _K}},\bar\beta_{{\xi _K}})}\omega\approx0$, or $\delta_{({\xi _K},\beta_{{\xi _K}},\bar\beta_{{\xi _K}})}h\approx0$ due to the assumption ${\cal L}_{\xi_K}\Omega\approx 0$ and \eqref{do}, we have
\begin{align}\label{beta2}
{{\cal L}_{{\xi _K}}}\omega  + \delta _{{\beta _{{\xi _K}}}}^{\rm{j}}\omega  + \delta _{{{\bar \beta }_{{\xi _K}}}}^{\rm{m}}\omega  = {{\cal L}_{{\xi _K}}}\omega  + {{\rm{d}}_\omega }{\beta _{{\xi _K}}} + \mu (1 + \sigma \alpha )\left[ {e,{{\bar \beta }_{{\xi _K}}}} \right] \approx 0,
\end{align}
which gives us
\begin{align}\label{relation b1b2}
&\left[ {e,{{\bar \beta }_{{\xi _K}}}} \right] \approx \frac{1}{{\mu (1 + \sigma \alpha )}}({{\cal L}_{{\xi _K}}}\omega  + {{\rm{d}}_\omega }{\beta _{{\xi _K}}}).
\end{align}

Note that without the additional part ${{\cal L}_{{\xi _K}}}\omega$ in maximally symmetric solution, one can solve ${{{\bar \beta }_{{\xi _K}}}}$ in terms of ${{{ \beta }_{{\xi _K}}}}$ by \eqref{relation b1b2}. To see this explicitly, we consider a zero-form $\phi^I$ and a one-form ${c_\mu}^I$, which satisfy ${\phi ^I} = \frac{1}{2}{\varepsilon ^I}_{JK}{e^{\mu J}}{c_\mu }^K$. If an anti-symmetric relationship ${c_\mu}^I=-{c^I}_\mu$ could be satisfied, then we have
\begin{align}
\begin{aligned}\label{eb2}
&{\varepsilon ^{IJK}}{e_{\mu J}}{\phi _K} =  - \frac{1}{2}(\delta _L^I\delta _M^{J} - \delta _M^I\delta _L^{J}){e_{\mu J}}{e^{\nu L}}{c_\nu }^{M}=  - \frac{1}{2}({c^I}_\mu  - {c_\mu }^I) = {c_\mu }^I.
 \end{aligned}
\end{align}
According to  \eqref{debeta}, this means $\bar\beta_{\xi_K}$ is zero. Since the term ${\left[ {e,{{\bar \beta }_{{\xi _K}}}} \right]}$ satisfies this anti-symmetric condition\footnote{For other black hole solutions (e.g., \cite{Alishahiha:2014dma}), extra care is required here since $\left[ {h,{{\bar \beta }_{{\xi _K}}}} \right]$ is not generally anti-symmetric.}, when the additional part in \eqref{eb2} is considered, we find that r.h.s. of \eqref{relation b1b2} in general fails to satisfy the anti-symmetric property. Thus, we need seek for another method  to solve $\beta_{\xi_K}$ and $\bar\beta_{\xi_K}$, respectively, as we will give in the next appendix.

\subsection{Seeking for the condition \eqref{51b} and constraint \eqref{parameters constraints} }
\label{appendix c2}
To ensure the generality, we now do not require ${\cal L}_{\xi_K} e$, ${\cal L}_{\xi_K} \Omega$ to vanish under on-shell condition (as the example of BTZ black hole in the main text). Actually, the condition \eqref{51b} is equivalent with $\delta_{(\xi_K,{\beta_{\xi_K}},{\bar\beta_{\xi_K}})}h\approx0$, when \eqref{51a} and \eqref{51b1} hold.

With the condition \eqref{51a} and \eqref{51b1}, we directly suppose
\begin{align}\label{Lie of Omega}
 \delta_{({\xi_K} ,\beta_{\xi_K},\bar\beta_{\xi_K})}h\approx 0,
\end{align}
which is equivalent with the requirement of \eqref{51b}.   This result can be obtained by \eqref{do} when $\delta_{(\xi_K,\beta_{\xi_K},\bar \beta_{\xi_K})}e\approx 0$ is assured by \eqref{51a} and $\delta_{(\xi_K,\beta_{\xi_K},\bar \beta_{\xi_K})}\Omega\approx 0$ by \eqref{51b1}, respectively. To this end, we directly assume $\lambda_{\xi_K}$ can be determined by
\begin{align}\label{main solution of beta}
 {{\cal K}_{{\xi _K}}}h = {{\cal L}_{{\xi _K}}}h + \left[ {h,{\lambda _{{\xi _K}}}} \right] \approx 0,
\end{align}
which is required by the maximally symmetric solution of MMG (see Section \ref{fg one} or \cite{MahdavianYekta:2015tmp}). Plugging the relation into the r.h.s. of \eqref{Lie of Omega} and requiring it to be zero\footnote{The logic here is to first assume it is zero, then verify whether we can uniquely find $\bar\beta_{\xi_K}$ that makes it zero.}, we have
\begin{align}
    \begin{aligned}\label{first order euqation of bar beta}
{{\rm{d}}_\omega }{{\bar \beta }_{{\xi _K}}} + \sigma \mu (1 + \sigma \alpha )\left[ {e,{{\bar \beta }_{{\xi _K}}}} \right] = 0
    \end{aligned}.
\end{align}

Now we turn to derive  the constraints \eqref{parameters constraints} adapted to the integrable  property with the assumption $1+\sigma\alpha\ne0$ in maximally symmetric solution. Combining  the first order equation of $\bar \beta_\xi$ \eqref{first order euqation of bar beta} and performing $\text{d}_\omega$ on the both side of it, we can obtain the curvature two-form $R(\omega)$, which should be compatible to the equation of motion \eqref{EOM}. We denote  this as the integrable condition of $\bar\beta_{\xi_K}$ ,  given by\footnote{This is similar to differential operators commute if their Lie bracket vanishes. This vanishing commutator is necessary and sufficient for the integrability of the associated vector fields via the Frobenius theorem \cite{nakahara2018geometry,Faddeev:1987ph}.}
\begin{align}\label{intecon}
\begin{aligned}
&{E^{in}} = \left[ {R(\omega ),{{\bar \beta }_{{\xi _K}}}} \right] + \sigma \mu (1 + \sigma \alpha )\left[ {{{\rm{d}}_\omega }e,{{\bar \beta }_{{\xi _K}}}} \right] - \sigma \mu (1 + \sigma \alpha )\left[ {e,{{\rm{d}}_\omega }{{\bar \beta }_{{\xi _K}}}} \right]\\
 &\approx - \mu (1 + \sigma \alpha )\left[ {\left[ {e \wedge h} \right],{{\bar \beta }_{{\xi _K}}}} \right] - \alpha \sigma \mu (1 + \sigma \alpha )\left[ {\left[ {e \wedge h} \right],{{\bar \beta }_{{\xi _K}}}} \right] - \sigma \mu (1 + \sigma \alpha )\left[ {e,{{\rm{d}}_\omega }{{\bar \beta }_{{\xi _K}}}} \right] = 0.
 \end{aligned}
\end{align}
Consider \eqref{first order euqation of bar beta} agian, \eqref{intecon} transfers into
\begin{align}
    \begin{aligned}
&{E^{in}} = - \mu (1 + \sigma \alpha )\left[ {\left[ {e \wedge h} \right],{{\bar \beta }_{{\xi _K}}}} \right] - \alpha \sigma \mu (1 + \sigma \alpha )\left[ {\left[ {e \wedge h} \right],{{\bar \beta }_{{\xi _K}}}} \right]\\
& + {\sigma ^2}{\mu ^2}{(1 + \sigma \alpha )^2}\left[ {e \wedge \left[ {e,{{\bar \beta }_{{\xi _K}}}} \right]} \right] = 0,
    \end{aligned}
\end{align}
from which we hope to constrain the parameters $\mu,\sigma,\alpha$ and $\Lambda_0$. Consider the solution of $h=C\mu e$, we can solve this integrable condition by
\begin{subequations}\label{conta 1}
\begin{align}
{\sigma ^2} = 2C.
\end{align}
\end{subequations}
With $C=\frac{{1 - \alpha {\ell ^2}{\Lambda _0}}}{{2{\ell ^2}{\mu ^2}{{(1 + \sigma \alpha )}^2}}}$,  we obtain the constraint \eqref{parameters constraints}, which are completely determined by the integrable condition \eqref{intecon} in solution $h=C\mu e$. Along with  \eqref{51a} and \eqref{51b1}, we can determine the formula of $\bar \beta_{\xi_K}$ with the additional condition  \eqref{51b}. Note that at the point $1+\sigma\alpha=0$, the integrability condition $E^{in}=0$ is automatically satisfied, which means that algebra \eqref{m algebras} is also satisfied. However, this case is beyond the scope of the example discussed in this paper.


\bibliography{ref}

\providecommand{\href}[2]{#2}\begingroup\raggedright\begin{thebibliography}{100}

\bibitem{Achucarro:1986uwr}
A.~Achucarro and P.~Townsend, ``{A Chern-Simons Action for Three-Dimensional
  anti-De Sitter Supergravity Theories},''
  \href{http://dx.doi.org/10.1016/0370-2693(86)90140-1}{{\em Phys. Lett. B}
  {\bfseries 180} (1986) 89}.

\bibitem{Witten:1988hc}
E.~Witten, ``{(2+1)-Dimensional Gravity as an Exactly Soluble System},'' {\em
  Nucl. Phys. B} {\bfseries 311} (1988) 46.

\bibitem{Deser:1982vy}
S.~Deser, R.~Jackiw, and S.~Templeton, ``{Three-Dimensional Massive Gauge
  Theories},'' \href{http://dx.doi.org/10.1103/PhysRevLett.48.975}{{\em Phys.
  Rev. Lett.} {\bfseries 48} (1982) 975}.

\bibitem{Li:2008dq}
W.~Li, W.~Song, and A.~Strominger, ``{Chiral Gravity in Three Dimensions},''
  \href{http://dx.doi.org/10.1088/1126-6708/2008/04/082}{{\em JHEP} {\bfseries
  04} (2008) 082}, \href{http://arxiv.org/abs/0801.4566}{{\ttfamily
  arXiv:0801.4566 [hep-th]}}.

\bibitem{Maloney:2007ud}
A.~Maloney and E.~Witten, ``{Quantum Gravity Partition Functions in Three
  Dimensions},'' \href{http://dx.doi.org/10.1007/JHEP02(2010)029}{{\em JHEP}
  {\bfseries 02} (2010) 029}, \href{http://arxiv.org/abs/0712.0155}{{\ttfamily
  arXiv:0712.0155 [hep-th]}}.

\bibitem{Carlip:2008jk}
S.~Carlip, S.~Deser, A.~Waldron, and D.~K. Wise, ``{Cosmological Topologically
  Massive Gravitons and Photons},''
  \href{http://dx.doi.org/10.1088/0264-9381/26/7/075008}{{\em Class. Quant.
  Grav.} {\bfseries 26} (2009) 075008},
  \href{http://arxiv.org/abs/0803.3998}{{\ttfamily arXiv:0803.3998 [hep-th]}}.

\bibitem{Grumiller:2008qz}
D.~Grumiller and N.~Johansson, ``{Instability in cosmological topologically
  massive gravity at the chiral point},''
  \href{http://dx.doi.org/10.1088/1126-6708/2008/07/134}{{\em JHEP} {\bfseries
  07} (2008) 134}, \href{http://arxiv.org/abs/0805.2610}{{\ttfamily
  arXiv:0805.2610 [hep-th]}}.

\bibitem{Skenderis:2009nt}
K.~Skenderis, M.~Taylor, and B.~C. van Rees, ``{Topologically Massive Gravity
  and the AdS/CFT Correspondence},''
  \href{http://dx.doi.org/10.1088/1126-6708/2009/09/045}{{\em JHEP} {\bfseries
  09} (2009) 045}, \href{http://arxiv.org/abs/0906.4926}{{\ttfamily
  arXiv:0906.4926 [hep-th]}}.

\bibitem{Blagojevic:2008bn}
M.~Blagojevic and B.~Cvetkovic, ``{Canonical structure of topologically massive
  gravity with a cosmological constant},''
  \href{http://dx.doi.org/10.1088/1126-6708/2009/05/073}{{\em JHEP} {\bfseries
  05} (2009) 073}, \href{http://arxiv.org/abs/0812.4742}{{\ttfamily
  arXiv:0812.4742 [gr-qc]}}.

\bibitem{Giribet:2008bw}
G.~Giribet, M.~Kleban, and M.~Porrati, ``{Topologically Massive Gravity at the
  Chiral Point is Not Chiral},''
  \href{http://dx.doi.org/10.1088/1126-6708/2008/10/045}{{\em JHEP} {\bfseries
  10} (2008) 045}, \href{http://arxiv.org/abs/0807.4703}{{\ttfamily
  arXiv:0807.4703 [hep-th]}}.

\bibitem{Grumiller:2008es}
D.~Grumiller and N.~Johansson, ``{Consistent boundary conditions for
  cosmological topologically massive gravity at the chiral point},''
  \href{http://dx.doi.org/10.1142/S0218271808014096}{{\em Int. J. Mod. Phys. D}
  {\bfseries 17} (2009) 2367--2372},
  \href{http://arxiv.org/abs/0808.2575}{{\ttfamily arXiv:0808.2575 [hep-th]}}.

\bibitem{Henneaux:2009pw}
M.~Henneaux, C.~Martinez, and R.~Troncoso, ``{Asymptotically anti-de Sitter
  spacetimes in topologically massive gravity},''
  \href{http://dx.doi.org/10.1103/PhysRevD.79.081502}{{\em Phys. Rev. D}
  {\bfseries 79} (2009) 081502},
  \href{http://arxiv.org/abs/0901.2874}{{\ttfamily arXiv:0901.2874 [hep-th]}}.

\bibitem{Bergshoeff:2014pca}
E.~Bergshoeff, O.~Hohm, W.~Merbis, A.~J. Routh, and P.~K. Townsend, ``{Minimal
  Massive 3D Gravity},''
  \href{http://dx.doi.org/10.1088/0264-9381/31/14/145008}{{\em Class. Quant.
  Grav.} {\bfseries 31} (2014) 145008},
  \href{http://arxiv.org/abs/1404.2867}{{\ttfamily arXiv:1404.2867 [hep-th]}}.

\bibitem{Arvanitakis:2014xna}
A.~S. Arvanitakis and P.~K. Townsend, ``{Minimal Massive 3D Gravity Unitarity
  Redux},'' \href{http://dx.doi.org/10.1088/0264-9381/32/8/085003}{{\em Class.
  Quant. Grav.} {\bfseries 32} no.~8, (2015) 085003},
  \href{http://arxiv.org/abs/1411.1970}{{\ttfamily arXiv:1411.1970 [hep-th]}}.

\bibitem{Alkac:2017vgg}
G.~Alkac, L.~Basanisi, E.~Kilicarslan, and B.~Tekin, ``{Unitarity Problems in
  3$D$ Gravity Theories},''
  \href{http://dx.doi.org/10.1103/PhysRevD.96.024010}{{\em Phys. Rev. D}
  {\bfseries 96} no.~2, (2017) 024010},
  \href{http://arxiv.org/abs/1703.03630}{{\ttfamily arXiv:1703.03630
  [hep-th]}}.

\bibitem{Bergshoeff:2009aq}
E.~A. Bergshoeff, O.~Hohm, and P.~K. Townsend, ``{More on Massive 3D
  Gravity},'' \href{http://dx.doi.org/10.1103/PhysRevD.79.124042}{{\em Phys.
  Rev. D} {\bfseries 79} (2009) 124042},
  \href{http://arxiv.org/abs/0905.1259}{{\ttfamily arXiv:0905.1259 [hep-th]}}.

\bibitem{Hohm:2012vh}
O.~Hohm, A.~Routh, P.~K. Townsend, and B.~Zhang, ``{On the Hamiltonian form of
  3D massive gravity},''
  \href{http://dx.doi.org/10.1103/PhysRevD.86.084035}{{\em Phys. Rev. D}
  {\bfseries 86} (2012) 084035},
  \href{http://arxiv.org/abs/1208.0038}{{\ttfamily arXiv:1208.0038 [hep-th]}}.

\bibitem{Bergshoeff:2014bia}
E.~A. Bergshoeff, O.~Hohm, W.~Merbis, A.~J. Routh, and P.~K. Townsend,
  ``{Chern-Simons-like Gravity Theories},''
  \href{http://dx.doi.org/10.1007/978-3-319-10070-8_7}{{\em Lect. Notes Phys.}
  {\bfseries 892} (2015) 181--201},
  \href{http://arxiv.org/abs/1402.1688}{{\ttfamily arXiv:1402.1688 [hep-th]}}.

\bibitem{MahdavianYekta:2015tmp}
D.~Mahdavian~Yekta, ``{Hamiltonian formalism of Minimal Massive Gravity},''
  \href{http://dx.doi.org/10.1103/PhysRevD.92.064044}{{\em Phys. Rev. D}
  {\bfseries 92} no.~6, (2015) 064044},
  \href{http://arxiv.org/abs/1503.08343}{{\ttfamily arXiv:1503.08343
  [hep-th]}}.

\bibitem{Alishahiha:2014dma}
M.~Alishahiha, M.~M. Qaemmaqami, A.~Naseh, and A.~Shirzad, ``{On 3D Minimal
  Massive Gravity},'' \href{http://dx.doi.org/10.1007/JHEP12(2014)033}{{\em
  JHEP} {\bfseries 12} (2014) 033},
  \href{http://arxiv.org/abs/1409.6146}{{\ttfamily arXiv:1409.6146 [hep-th]}}.

\bibitem{Hajian:2015xlp}
K.~Hajian and M.~M. Sheikh-Jabbari, ``{Solution Phase Space and Conserved
  Charges: A General Formulation for Charges Associated with Exact
  Symmetries},'' \href{http://dx.doi.org/10.1103/PhysRevD.93.044074}{{\em Phys.
  Rev. D} {\bfseries 93} no.~4, (2016) 044074},
  \href{http://arxiv.org/abs/1512.05584}{{\ttfamily arXiv:1512.05584
  [hep-th]}}.

\bibitem{Tavlayan:2024zbl}
A.~Tavlayan and B.~Tekin, ``{Refined thermodynamics of black holes with proper
  conserved charges},''
  \href{http://dx.doi.org/10.1103/PhysRevD.110.084049}{{\em Phys. Rev. D}
  {\bfseries 110} no.~8, (2024) 084049},
  \href{http://arxiv.org/abs/2407.04002}{{\ttfamily arXiv:2407.04002 [gr-qc]}}.

\bibitem{Deger:2023eah}
N.~S. Deger, M.~Geiller, J.~Rosseel, and H.~Samtleben, ``{Minimal massive
  supergravity and new theories of massive gravity},''
  \href{http://dx.doi.org/10.1103/PhysRevD.109.086014}{{\em Phys. Rev. D}
  {\bfseries 109} no.~8, (2024) 086014},
  \href{http://arxiv.org/abs/2312.12387}{{\ttfamily arXiv:2312.12387
  [hep-th]}}.

\bibitem{Xiao:2023lap}
Y.~Xiao, Y.~Tian, and Y.-X. Liu, ``{Extended Black Hole Thermodynamics from
  Extended Iyer-Wald Formalism},''
  \href{http://dx.doi.org/10.1103/PhysRevLett.132.021401}{{\em Phys. Rev.
  Lett.} {\bfseries 132} no.~2, (2024) 021401},
  \href{http://arxiv.org/abs/2308.12630}{{\ttfamily arXiv:2308.12630 [gr-qc]}}.

\bibitem{Hajian:2023bhq}
K.~Hajian and B.~Tekin, ``{Coupling Constants as Conserved Charges in Black
  Hole Thermodynamics},''
  \href{http://dx.doi.org/10.1103/PhysRevLett.132.191401}{{\em Phys. Rev.
  Lett.} {\bfseries 132} no.~19, (2024) 191401},
  \href{http://arxiv.org/abs/2309.07634}{{\ttfamily arXiv:2309.07634 [gr-qc]}}.

\bibitem{Geiller:2020edh}
M.~Geiller, C.~Goeller, and N.~Merino, ``{Most general theory of 3d gravity:
  Covariant phase space, dual diffeomorphisms, and more},''
  \href{http://dx.doi.org/10.1007/JHEP02(2021)120}{{\em JHEP} {\bfseries 02}
  (2021) 120}, \href{http://arxiv.org/abs/2011.09873}{{\ttfamily
  arXiv:2011.09873 [hep-th]}}.

\bibitem{Banerjee:2012jn}
R.~Banerjee and D.~Roy, ``{Trivial symmetries in a 3D topological torsion model
  of gravity},'' \href{http://dx.doi.org/10.1088/1742-6596/405/1/012028}{{\em
  J. Phys. Conf. Ser.} {\bfseries 405} (2012) 012028},
  \href{http://arxiv.org/abs/1212.4238}{{\ttfamily arXiv:1212.4238 [gr-qc]}}.

\bibitem{Banerjee:2009vf}
R.~Banerjee, S.~Gangopadhyay, P.~Mukherjee, and D.~Roy, ``{Symmetries of
  topological gravity with torsion in the hamiltonian and lagrangian
  formalisms},'' \href{http://dx.doi.org/10.1007/JHEP02(2010)075}{{\em JHEP}
  {\bfseries 02} (2010) 075}, \href{http://arxiv.org/abs/0912.1472}{{\ttfamily
  arXiv:0912.1472 [gr-qc]}}.

\bibitem{Banerjee:2011cu}
R.~Banerjee and D.~Roy, ``{Poincare gauge symmetries, hamiltonian symmetries
  and trivial gauge transformations},''
  \href{http://dx.doi.org/10.1103/PhysRevD.84.124034}{{\em Phys. Rev. D}
  {\bfseries 84} (2011) 124034},
  \href{http://arxiv.org/abs/1110.1720}{{\ttfamily arXiv:1110.1720 [gr-qc]}}.

\bibitem{Yilmaz:2007ij}
N.~T. Yilmaz, ``{On the symmetric space sigma-model kinematics},''
  \href{http://dx.doi.org/10.1142/S0217751X07036324}{{\em Int. J. Mod. Phys. A}
  {\bfseries 22} (2007) 2683--2695},
  \href{http://arxiv.org/abs/0707.2150}{{\ttfamily arXiv:0707.2150 [hep-th]}}.

\bibitem{Keurentjes:2002xc}
A.~Keurentjes, ``{The Group theory of oxidation},''
  \href{http://dx.doi.org/10.1016/S0550-3213(03)00178-0}{{\em Nucl. Phys. B}
  {\bfseries 658} (2003) 303--347},
  \href{http://arxiv.org/abs/hep-th/0210178}{{\ttfamily arXiv:hep-th/0210178}}.

\bibitem{DePaoli:2018erh}
E.~De~Paoli and S.~Speziale, ``{A gauge-invariant symplectic potential for
  tetrad general relativity},''
  \href{http://dx.doi.org/10.1007/JHEP07(2018)040}{{\em JHEP} {\bfseries 07}
  (2018) 040}, \href{http://arxiv.org/abs/1804.09685}{{\ttfamily
  arXiv:1804.09685 [gr-qc]}}.

\bibitem{Jubb:2016qzt}
I.~Jubb, J.~Samuel, R.~Sorkin, and S.~Surya, ``{Boundary and Corner Terms in
  the Action for General Relativity},''
  \href{http://dx.doi.org/10.1088/1361-6382/aa6014}{{\em Class. Quant. Grav.}
  {\bfseries 34} no.~6, (2017) 065006},
  \href{http://arxiv.org/abs/1612.00149}{{\ttfamily arXiv:1612.00149 [gr-qc]}}.

\bibitem{Wieland:2017zkf}
W.~Wieland, ``{New boundary variables for classical and quantum gravity on a
  null surface},'' \href{http://dx.doi.org/10.1088/1361-6382/aa8d06}{{\em
  Class. Quant. Grav.} {\bfseries 34} no.~21, (2017) 215008},
  \href{http://arxiv.org/abs/1704.07391}{{\ttfamily arXiv:1704.07391 [gr-qc]}}.

\bibitem{Oliveri:2019gvm}
R.~Oliveri and S.~Speziale, ``{Boundary effects in General Relativity with
  tetrad variables},'' \href{http://dx.doi.org/10.1007/s10714-020-02733-8}{{\em
  Gen. Rel. Grav.} {\bfseries 52} no.~8, (2020) 83},
  \href{http://arxiv.org/abs/1912.01016}{{\ttfamily arXiv:1912.01016 [gr-qc]}}.

\bibitem{Frodden:2019ylc}
E.~Frodden and D.~Hidalgo, ``{Surface Charges Toolkit for Gravity},''
  \href{http://dx.doi.org/10.1142/S0218271820500406}{{\em Int. J. Mod. Phys. D}
  {\bfseries 29} no.~06, (2020) 2050040},
  \href{http://arxiv.org/abs/1911.07264}{{\ttfamily arXiv:1911.07264
  [hep-th]}}.

\bibitem{Kiran:2014dfa}
K.~S. Kiran, C.~Krishnan, and A.~Raju, ``{3D gravity, Chern\textendash{}Simons
  and higher spins: A mini introduction},''
  \href{http://dx.doi.org/10.1142/S0217732315300232}{{\em Mod. Phys. Lett. A}
  {\bfseries 30} no.~32, (2015) 1530023},
  \href{http://arxiv.org/abs/1412.5053}{{\ttfamily arXiv:1412.5053 [hep-th]}}.

\bibitem{Zanelli:2005sa}
J.~Zanelli, ``{Lecture notes on Chern-Simons (super-)gravities. Second edition
  (February 2008)},'' in {\em {7th Mexican Workshop on Particles and Fields}}.
\newblock 2, 2005.
\newblock \href{http://arxiv.org/abs/hep-th/0502193}{{\ttfamily
  arXiv:hep-th/0502193}}.

\bibitem{Wald:1993nt}
R.~M. Wald, ``{Black hole entropy is the Noether charge},''
  \href{http://dx.doi.org/10.1103/PhysRevD.48.R3427}{{\em Phys. Rev. D}
  {\bfseries 48} no.~8, (1993) R3427--R3431},
  \href{http://arxiv.org/abs/gr-qc/9307038}{{\ttfamily arXiv:gr-qc/9307038}}.

\bibitem{Jacobson:2015uqa}
T.~Jacobson and A.~Mohd, ``{Black hole entropy and Lorentz-diffeomorphism
  Noether charge},'' \href{http://dx.doi.org/10.1103/PhysRevD.92.124010}{{\em
  Phys. Rev. D} {\bfseries 92} (2015) 124010},
  \href{http://arxiv.org/abs/1507.01054}{{\ttfamily arXiv:1507.01054 [gr-qc]}}.

\bibitem{Cacciatori:2005wz}
S.~L. Cacciatori, M.~M. Caldarelli, A.~Giacomini, D.~Klemm, and D.~S. Mansi,
  ``{Chern-Simons formulation of three-dimensional gravity with torsion and
  nonmetricity},'' \href{http://dx.doi.org/10.1016/j.geomphys.2006.01.006}{{\em
  J. Geom. Phys.} {\bfseries 56} (2006) 2523--2543},
  \href{http://arxiv.org/abs/hep-th/0507200}{{\ttfamily arXiv:hep-th/0507200}}.

\bibitem{Giacomini:2006dr}
A.~Giacomini, R.~Troncoso, and S.~Willison, ``{Three-dimensional supergravity
  reloaded},'' \href{http://dx.doi.org/10.1088/0264-9381/24/11/005}{{\em Class.
  Quant. Grav.} {\bfseries 24} (2007) 2845--2860},
  \href{http://arxiv.org/abs/hep-th/0610077}{{\ttfamily arXiv:hep-th/0610077}}.

\bibitem{Charyyev:2017uuu}
J.~Charyyev and N.~S. Deger, ``{Homogeneous Solutions of Minimal Massive 3D
  Gravity},'' \href{http://dx.doi.org/10.1103/PhysRevD.96.026024}{{\em Phys.
  Rev. D} {\bfseries 96} no.~2, (2017) 026024},
  \href{http://arxiv.org/abs/1703.06871}{{\ttfamily arXiv:1703.06871
  [hep-th]}}.

\bibitem{Alishahiha:2015whv}
M.~Alishahiha, M.~M. Qaemmaqami, A.~Naseh, and A.~Shirzad, ``{Holographic
  Renormalization of 3D Minimal Massive Gravity},''
  \href{http://dx.doi.org/10.1007/JHEP01(2016)106}{{\em JHEP} {\bfseries 01}
  (2016) 106}, \href{http://arxiv.org/abs/1511.06194}{{\ttfamily
  arXiv:1511.06194 [hep-th]}}.

\bibitem{Wald:1999wa}
R.~M. Wald and A.~Zoupas, ``{A General definition of 'conserved quantities' in
  general relativity and other theories of gravity},''
  \href{http://dx.doi.org/10.1103/PhysRevD.61.084027}{{\em Phys. Rev. D}
  {\bfseries 61} (2000) 084027},
  \href{http://arxiv.org/abs/gr-qc/9911095}{{\ttfamily arXiv:gr-qc/9911095}}.

\bibitem{Poisson:2009pwt}
E.~Poisson, \href{http://dx.doi.org/10.1017/CBO9780511606601}{{\em {A
  Relativist's Toolkit: The Mathematics of Black-Hole Mechanics}}}.
\newblock Cambridge University Press, 12, 2009.

\bibitem{Setare:2015nla}
M.~R. Setare and H.~Adami, ``{Black hole entropy in the
  Chern\textendash{}Simons-like theories of gravity and Lorentz-diffeomorphism
  Noether charge},''
  \href{http://dx.doi.org/10.1016/j.nuclphysb.2015.11.018}{{\em Nucl. Phys. B}
  {\bfseries 902} (2016) 115--123},
  \href{http://arxiv.org/abs/1509.05972}{{\ttfamily arXiv:1509.05972
  [hep-th]}}.

\bibitem{Adami:2015onz}
H.~Adami and M.~R. Setare, ``{Quasi-local conserved charges in
  Lorenz\textendash{}diffeomorphism covariant theory of gravity},''
  \href{http://dx.doi.org/10.1140/epjc/s10052-016-4032-x}{{\em Eur. Phys. J. C}
  {\bfseries 76} no.~4, (2016) 187},
  \href{http://arxiv.org/abs/1511.00527}{{\ttfamily arXiv:1511.00527 [gr-qc]}}.

\bibitem{Setare:2015gss}
M.~R. Setare and H.~Adami, ``{Lorentz-diffeomorphism quasi-local conserved
  charges and Virasoro algebra in Chern\textendash{}Simons-like theories of
  gravity},'' \href{http://dx.doi.org/10.1016/j.nuclphysb.2016.05.020}{{\em
  Nucl. Phys. B} {\bfseries 909} (2016) 345--359},
  \href{http://arxiv.org/abs/1511.01070}{{\ttfamily arXiv:1511.01070
  [hep-th]}}.

\bibitem{Setare:2017wuj}
M.~R. Setare and H.~Adami, ``{General formulae for conserved charges and black
  hole entropy in Chern-Simons-like theories of gravity},''
  \href{http://dx.doi.org/10.1103/PhysRevD.96.104019}{{\em Phys. Rev. D}
  {\bfseries 96} no.~10, (2017) 104019},
  \href{http://arxiv.org/abs/1708.08767}{{\ttfamily arXiv:1708.08767
  [hep-th]}}.

\bibitem{Adami:2017phg}
H.~Adami, M.~R. Setare, T.~C. Sisman, and B.~Tekin, ``{Conserved Charges in
  Extended Theories of Gravity},''
  \href{http://dx.doi.org/10.1016/j.physrep.2019.08.003}{{\em Phys. Rept.}
  {\bfseries 834} (2019) 1}, \href{http://arxiv.org/abs/1710.07252}{{\ttfamily
  arXiv:1710.07252 [hep-th]}}.

\bibitem{Hollands:2024vbe}
S.~Hollands, R.~M. Wald, and V.~G. Zhang, ``{Entropy of dynamical black
  holes},'' \href{http://dx.doi.org/10.1103/PhysRevD.110.024070}{{\em Phys.
  Rev. D} {\bfseries 110} no.~2, (2024) 024070},
  \href{http://arxiv.org/abs/2402.00818}{{\ttfamily arXiv:2402.00818
  [hep-th]}}.

\bibitem{Kong:2024sqc}
D.~Kong, Y.~Tian, H.~Zhang, and J.~Zhao, ``{Dynamical black hole entropy beyond
  general relativity from the Einstein frame},''
  \href{http://dx.doi.org/10.1103/PhysRevD.111.084005}{{\em Phys. Rev. D}
  {\bfseries 111} no.~8, (2025) 084005},
  \href{http://arxiv.org/abs/2412.00647}{{\ttfamily arXiv:2412.00647
  [hep-th]}}.

\bibitem{Jiang:2018sqj}
J.~Jiang and H.~Zhang, ``{Surface term, corner term, and action growth in
  $F(R_{abcd})$ gravity theory},''
  \href{http://dx.doi.org/10.1103/PhysRevD.99.086005}{{\em Phys. Rev. D}
  {\bfseries 99} no.~8, (2019) 086005},
  \href{http://arxiv.org/abs/1806.10312}{{\ttfamily arXiv:1806.10312
  [hep-th]}}.

\bibitem{Guo:2024oey}
W.~Guo, X.~Guo, M.~Li, Z.~Mou, and H.~Zhang, ``{Equivalence of Noether charge
  and Hilbert action boundary term formulas for the black hole entropy in
  F(Rabcd) gravity theory},''
  \href{http://dx.doi.org/10.1103/PhysRevD.110.064071}{{\em Phys. Rev. D}
  {\bfseries 110} no.~6, (2024) 064071},
  \href{http://arxiv.org/abs/2406.15138}{{\ttfamily arXiv:2406.15138
  [hep-th]}}.

\bibitem{Boulanger:2023tvt}
N.~Boulanger, A.~Campoleoni, V.~Lekeu, and E.~Skvortsov, ``{Strange higher-spin
  topological systems in 3D},''
  \href{http://dx.doi.org/10.1007/JHEP05(2024)109}{{\em JHEP} {\bfseries 05}
  (2024) 109}, \href{http://arxiv.org/abs/2312.03382}{{\ttfamily
  arXiv:2312.03382 [hep-th]}}.

\bibitem{Merbis:2014vja}
W.~Merbis, ``{Chern-Simons-like Theories of Gravity},'' other thesis, 11, 2014.

\bibitem{Bergshoeff:2009zz}
E.~A. Bergshoeff, O.~Hohm, and P.~K. Townsend,
  \href{http://dx.doi.org/10.1142/9789814374552_0470}{``{New massive
  gravity},''} in {\em {12th Marcel Grossmann Meeting on General Relativity}},
  pp.~2329--2331.
\newblock 7, 2009.

\bibitem{Bergshoeff:2013xma}
E.~A. Bergshoeff, S.~de~Haan, O.~Hohm, W.~Merbis, and P.~K. Townsend,
  ``{Zwei-Dreibein Gravity: A Two-Frame-Field Model of 3D Massive Gravity},''
  \href{http://dx.doi.org/10.1103/PhysRevLett.111.111102}{{\em Phys. Rev.
  Lett.} {\bfseries 111} no.~11, (2013) 111102},
  \href{http://arxiv.org/abs/1307.2774}{{\ttfamily arXiv:1307.2774 [hep-th]}}.
  [Erratum: Phys.Rev.Lett. 111, 259902 (2013)].

\bibitem{Blagojevic:2010ir}
M.~Blagojevic and B.~Cvetkovic, ``{Hamiltonian analysis of BHT massive
  gravity},'' \href{http://dx.doi.org/10.1007/JHEP01(2011)082}{{\em JHEP}
  {\bfseries 01} (2011) 082}, \href{http://arxiv.org/abs/1010.2596}{{\ttfamily
  arXiv:1010.2596 [gr-qc]}}.

\bibitem{Tekin:2014jna}
B.~Tekin, ``{Minimal Massive Gravity: Conserved Charges, Excitations and the
  Chiral Gravity Limit},''
  \href{http://dx.doi.org/10.1103/PhysRevD.90.081701}{{\em Phys. Rev. D}
  {\bfseries 90} no.~8, (2014) 081701},
  \href{http://arxiv.org/abs/1409.5358}{{\ttfamily arXiv:1409.5358 [hep-th]}}.

\bibitem{Zanelli:2012px}
J.~Zanelli, ``{Chern-Simons Forms in Gravitation Theories},''
  \href{http://dx.doi.org/10.1088/0264-9381/29/13/133001}{{\em Class. Quant.
  Grav.} {\bfseries 29} (2012) 133001},
  \href{http://arxiv.org/abs/1208.3353}{{\ttfamily arXiv:1208.3353 [hep-th]}}.

\bibitem{Barnich:2010eb}
G.~Barnich and C.~Troessaert, ``{Aspects of the BMS/CFT correspondence},''
  \href{http://dx.doi.org/10.1007/JHEP05(2010)062}{{\em JHEP} {\bfseries 05}
  (2010) 062}, \href{http://arxiv.org/abs/1001.1541}{{\ttfamily arXiv:1001.1541
  [hep-th]}}.

\bibitem{Compere:2018ylh}
G.~Comp\`ere, A.~Fiorucci, and R.~Ruzziconi, ``{Superboost transitions,
  refraction memory and super-Lorentz charge algebra},''
  \href{http://dx.doi.org/10.1007/JHEP11(2018)200}{{\em JHEP} {\bfseries 11}
  (2018) 200}, \href{http://arxiv.org/abs/1810.00377}{{\ttfamily
  arXiv:1810.00377 [hep-th]}}. [Erratum: JHEP 04, 172 (2020)].

\bibitem{Compere:2020lrt}
G.~Comp\`ere, A.~Fiorucci, and R.~Ruzziconi, ``{The $\Lambda$-BMS$_4$ charge
  algebra},'' \href{http://dx.doi.org/10.1007/JHEP10(2020)205}{{\em JHEP}
  {\bfseries 10} (2020) 205}, \href{http://arxiv.org/abs/2004.10769}{{\ttfamily
  arXiv:2004.10769 [hep-th]}}.

\bibitem{Setare:2015pva}
M.~R. Setare and H.~Adami, ``{Entropy formula of black holes in minimal massive
  gravity and its application for BTZ black holes},''
  \href{http://dx.doi.org/10.1103/PhysRevD.91.104039}{{\em Phys. Rev. D}
  {\bfseries 91} no.~10, (2015) 104039},
  \href{http://arxiv.org/abs/1501.00920}{{\ttfamily arXiv:1501.00920
  [hep-th]}}.

\bibitem{Lee:1990nz}
J.~Lee and R.~M. Wald, ``{Local symmetries and constraints},''
  \href{http://dx.doi.org/10.1063/1.528801}{{\em J. Math. Phys.} {\bfseries 31}
  (1990) 725--743}.

\bibitem{Harlow:2019yfa}
D.~Harlow and J.-Q. Wu, ``{Covariant phase space with boundaries},''
  \href{http://dx.doi.org/10.1007/JHEP10(2020)146}{{\em JHEP} {\bfseries 10}
  (2020) 146}, \href{http://arxiv.org/abs/1906.08616}{{\ttfamily
  arXiv:1906.08616 [hep-th]}}.

\bibitem{Prabhu:2015vua}
K.~Prabhu, ``{The First Law of Black Hole Mechanics for Fields with Internal
  Gauge Freedom},'' \href{http://dx.doi.org/10.1088/1361-6382/aa536b}{{\em
  Class. Quant. Grav.} {\bfseries 34} no.~3, (2017) 035011},
  \href{http://arxiv.org/abs/1511.00388}{{\ttfamily arXiv:1511.00388 [gr-qc]}}.

\bibitem{JACKIW1980257}
R.~Jackiw and N.~Manton, ``Symmetries and conservation laws in gauge
  theories,''
  \href{http://dx.doi.org/https://doi.org/10.1016/0003-4916(80)90098-6}{{\em
  Annals of Physics} {\bfseries 127} no.~2, (1980) 257--273}.
  \url{https://www.sciencedirect.com/science/article/pii/0003491680900986}.

\bibitem{Obukhov:2006ge}
Y.~N. Obukhov and G.~F. Rubilar, ``{Invariant conserved currents in gravity
  theories with local Lorentz and diffeomorphism symmetry},''
  \href{http://dx.doi.org/10.1103/PhysRevD.74.064002}{{\em Phys. Rev. D}
  {\bfseries 74} (2006) 064002},
  \href{http://arxiv.org/abs/gr-qc/0608064}{{\ttfamily arXiv:gr-qc/0608064}}.

\bibitem{fatibene2009general}
L.~Fatibene and M.~Francaviglia, ``General theory of lie derivatives for
  lorentz tensors,'' {\em arXiv preprint arXiv:0904.0258} (2009) .

\bibitem{Iyer:1994ys}
V.~Iyer and R.~M. Wald, ``{Some properties of Noether charge and a proposal for
  dynamical black hole entropy},''
  \href{http://dx.doi.org/10.1103/PhysRevD.50.846}{{\em Phys. Rev. D}
  {\bfseries 50} (1994) 846--864},
  \href{http://arxiv.org/abs/gr-qc/9403028}{{\ttfamily arXiv:gr-qc/9403028}}.

\bibitem{Ghodrati:2016vvf}
M.~Ghodrati, K.~Hajian, and M.~R. Setare, ``{Revisiting Conserved Charges in
  Higher Curvature Gravitational Theories},''
  \href{http://dx.doi.org/10.1140/epjc/s10052-016-4550-6}{{\em Eur. Phys. J. C}
  {\bfseries 76} no.~12, (2016) 701},
  \href{http://arxiv.org/abs/1606.04353}{{\ttfamily arXiv:1606.04353
  [hep-th]}}.

\bibitem{Kastor:2009wy}
D.~Kastor, S.~Ray, and J.~Traschen, ``{Enthalpy and the Mechanics of AdS Black
  Holes},'' \href{http://dx.doi.org/10.1088/0264-9381/26/19/195011}{{\em Class.
  Quant. Grav.} {\bfseries 26} (2009) 195011},
  \href{http://arxiv.org/abs/0904.2765}{{\ttfamily arXiv:0904.2765 [hep-th]}}.

\bibitem{Kraus:2005zm}
P.~Kraus and F.~Larsen, ``{Holographic gravitational anomalies},''
  \href{http://dx.doi.org/10.1088/1126-6708/2006/01/022}{{\em JHEP} {\bfseries
  01} (2006) 022}, \href{http://arxiv.org/abs/hep-th/0508218}{{\ttfamily
  arXiv:hep-th/0508218}}.

\bibitem{Miskovic:2006tm}
O.~Miskovic and R.~Olea, ``{On boundary conditions in three-dimensional AdS
  gravity},'' \href{http://dx.doi.org/10.1016/j.physletb.2006.07.045}{{\em
  Phys. Lett. B} {\bfseries 640} (2006) 101--107},
  \href{http://arxiv.org/abs/hep-th/0603092}{{\ttfamily arXiv:hep-th/0603092}}.

\bibitem{Grumiller:2015xaa}
D.~Grumiller and W.~Merbis, ``{Free energy of topologically massive gravity and
  flat space holography},''
  \href{http://dx.doi.org/10.1007/978-3-319-94256-8_10}{{\em Springer Proc.
  Phys.} {\bfseries 208} (2018) 95--103},
  \href{http://arxiv.org/abs/1509.08505}{{\ttfamily arXiv:1509.08505
  [hep-th]}}.

\bibitem{MahdavianYekta:2016kqh}
D.~Mahdavian~Yekta, ``{Entropy product of rotating black holes in
  three-dimensions},'' \href{http://dx.doi.org/10.1103/PhysRevD.95.064027}{{\em
  Phys. Rev. D} {\bfseries 95} no.~6, (2017) 064027},
  \href{http://arxiv.org/abs/1612.01135}{{\ttfamily arXiv:1612.01135
  [hep-th]}}.

\bibitem{Wei:2018sqy}
C.-H. Wei and B.~Ning, ``{Quasi-local Energy in 3D Gravity with Torsion},''
  \href{http://arxiv.org/abs/1807.08736}{{\ttfamily arXiv:1807.08736
  [hep-th]}}.

\bibitem{Clement:2003sr}
G.~Clement, ``{Black hole mass and angular momentum in 2+1 gravity},''
  \href{http://dx.doi.org/10.1103/PhysRevD.68.024032}{{\em Phys. Rev. D}
  {\bfseries 68} (2003) 024032},
  \href{http://arxiv.org/abs/gr-qc/0301129}{{\ttfamily arXiv:gr-qc/0301129}}.

\bibitem{Moussa:2003fc}
K.~A. Moussa, G.~Clement, and C.~Leygnac, ``{The Black holes of topologically
  massive gravity},'' \href{http://dx.doi.org/10.1088/0264-9381/20/24/L01}{{\em
  Class. Quant. Grav.} {\bfseries 20} (2003) L277--L283},
  \href{http://arxiv.org/abs/gr-qc/0303042}{{\ttfamily arXiv:gr-qc/0303042}}.

\bibitem{deRham:2010ik}
C.~de~Rham and G.~Gabadadze, ``{Generalization of the Fierz-Pauli Action},''
  \href{http://dx.doi.org/10.1103/PhysRevD.82.044020}{{\em Phys. Rev. D}
  {\bfseries 82} (2010) 044020},
  \href{http://arxiv.org/abs/1007.0443}{{\ttfamily arXiv:1007.0443 [hep-th]}}.

\bibitem{deRham:2010kj}
C.~de~Rham, G.~Gabadadze, and A.~J. Tolley, ``{Resummation of Massive
  Gravity},'' \href{http://dx.doi.org/10.1103/PhysRevLett.106.231101}{{\em
  Phys. Rev. Lett.} {\bfseries 106} (2011) 231101},
  \href{http://arxiv.org/abs/1011.1232}{{\ttfamily arXiv:1011.1232 [hep-th]}}.

\bibitem{Sinha:2010ai}
A.~Sinha, ``{On the new massive gravity and AdS/CFT},''
  \href{http://dx.doi.org/10.1007/JHEP06(2010)061}{{\em JHEP} {\bfseries 06}
  (2010) 061}, \href{http://arxiv.org/abs/1003.0683}{{\ttfamily arXiv:1003.0683
  [hep-th]}}.

\bibitem{Gallagher:2023ghl}
P.~Gallagher, T.~S. Koivisto, L.~Marzola, L.~Varrin, and T.~Zlosnik,
  ``{Consistent first-order action functional for gauge theories},''
  \href{http://dx.doi.org/10.1103/PhysRevD.109.L061503}{{\em Phys. Rev. D}
  {\bfseries 109} no.~6, (2024) L061503},
  \href{http://arxiv.org/abs/2311.07464}{{\ttfamily arXiv:2311.07464
  [hep-th]}}.

\bibitem{Andrianopoli:2023dfm}
L.~Andrianopoli, B.~L. Cerchiai, R.~Noris, L.~Ravera, M.~Trigiante, and
  J.~Zanelli, ``{New torsional deformations of locally AdS3 space},''
  \href{http://dx.doi.org/10.1103/PhysRevD.108.044011}{{\em Phys. Rev. D}
  {\bfseries 108} no.~4, (2023) 044011},
  \href{http://arxiv.org/abs/2305.17168}{{\ttfamily arXiv:2305.17168
  [hep-th]}}.

\bibitem{Borsten:2024pfz}
L.~Borsten, D.~Kanakaris, and H.~Kim, ``{Three-dimensional SL(2,R) Yang-Mills
  theory is equivalent to three-dimensional gravity with background sources},''
  \href{http://dx.doi.org/10.1103/PhysRevD.111.025005}{{\em Phys. Rev. D}
  {\bfseries 111} no.~2, (2025) 025005},
  \href{http://arxiv.org/abs/2408.14228}{{\ttfamily arXiv:2408.14228
  [hep-th]}}.

\bibitem{Grumiller:2016pqb}
D.~Grumiller and M.~Riegler, ``{Most general AdS$_{3}$ boundary conditions},''
  \href{http://dx.doi.org/10.1007/JHEP10(2016)023}{{\em JHEP} {\bfseries 10}
  (2016) 023}, \href{http://arxiv.org/abs/1608.01308}{{\ttfamily
  arXiv:1608.01308 [hep-th]}}.

\bibitem{Fiorucci:2021pha}
A.~Fiorucci, {\em {Leaky covariant phase spaces: Theory and application to
  $\Lambda$-BMS symmetry}}.
\newblock PhD thesis, Brussels U., Intl. Solvay Inst., Brussels, 2021.
\newblock \href{http://arxiv.org/abs/2112.07666}{{\ttfamily arXiv:2112.07666
  [hep-th]}}.

\bibitem{McNees:2024iyu}
R.~McNees and C.~Zwikel, ``{The symplectic potential for leaky boundaries},''
  \href{http://dx.doi.org/10.1007/JHEP01(2025)049}{{\em JHEP} {\bfseries 01}
  (2025) 049}, \href{http://arxiv.org/abs/2408.13203}{{\ttfamily
  arXiv:2408.13203 [hep-th]}}.

\bibitem{Grumiller:2017qao}
D.~Grumiller, R.~McNees, J.~Salzer, C.~Valc{\'a}rcel, and D.~Vassilevich,
  ``{Menagerie of AdS$_{2}$ boundary conditions},''
  \href{http://dx.doi.org/10.1007/JHEP10(2017)203}{{\em JHEP} {\bfseries 10}
  (2017) 203}, \href{http://arxiv.org/abs/1708.08471}{{\ttfamily
  arXiv:1708.08471 [hep-th]}}.

\bibitem{Gonzalez:2018enk}
H.~A. Gonz{\'a}lez, D.~Grumiller, and J.~Salzer, ``{Towards a bulk description
  of higher spin SYK},'' \href{http://dx.doi.org/10.1007/JHEP05(2018)083}{{\em
  JHEP} {\bfseries 05} (2018) 083},
  \href{http://arxiv.org/abs/1802.01562}{{\ttfamily arXiv:1802.01562
  [hep-th]}}.

\bibitem{Freidel:2020svx}
L.~Freidel, M.~Geiller, and D.~Pranzetti, ``{Edge modes of gravity. Part II.
  Corner metric and Lorentz charges},''
  \href{http://dx.doi.org/10.1007/JHEP11(2020)027}{{\em JHEP} {\bfseries 11}
  (2020) 027}, \href{http://arxiv.org/abs/2007.03563}{{\ttfamily
  arXiv:2007.03563 [hep-th]}}.

\bibitem{Freidel:2020ayo}
L.~Freidel, M.~Geiller, and D.~Pranzetti, ``{Edge modes of gravity. Part III.
  Corner simplicity constraints},''
  \href{http://dx.doi.org/10.1007/JHEP01(2021)100}{{\em JHEP} {\bfseries 01}
  (2021) 100}, \href{http://arxiv.org/abs/2007.12635}{{\ttfamily
  arXiv:2007.12635 [hep-th]}}.

\bibitem{Geiller:2021vpg}
M.~Geiller, C.~Goeller, and C.~Zwikel, ``{3d gravity in Bondi-Weyl gauge:
  charges, corners, and integrability},''
  \href{http://dx.doi.org/10.1007/JHEP09(2021)029}{{\em JHEP} {\bfseries 09}
  (2021) 029}, \href{http://arxiv.org/abs/2107.01073}{{\ttfamily
  arXiv:2107.01073 [hep-th]}}.

\bibitem{Ciambelli:2024vhy}
L.~Ciambelli and M.~Geiller, ``{Field-dependent diffeomorphisms and the
  transformation of surface charges between gauges},''
  \href{http://dx.doi.org/10.1007/JHEP05(2025)022}{{\em JHEP} {\bfseries 05}
  (2025) 022}, \href{http://arxiv.org/abs/2412.14992}{{\ttfamily
  arXiv:2412.14992 [hep-th]}}.

\bibitem{Poole:2018koa}
A.~Poole, K.~Skenderis, and M.~Taylor, ``{(A)dS$\mathbf{_4}$ in Bondi gauge},''
  \href{http://dx.doi.org/10.1088/1361-6382/ab117c}{{\em Class. Quant. Grav.}
  {\bfseries 36} no.~9, (2019) 095005},
  \href{http://arxiv.org/abs/1812.05369}{{\ttfamily arXiv:1812.05369
  [hep-th]}}.

\bibitem{Compere:2023ktn}
G.~Comp{\`e}re, S.~J. Hoque, and E.~{\c{S}}. Kutluk, ``{Quadrupolar radiation
  in de Sitter: displacement memory and Bondi metric},''
  \href{http://dx.doi.org/10.1088/1361-6382/ad5826}{{\em Class. Quant. Grav.}
  {\bfseries 41} no.~15, (2024) 155006},
  \href{http://arxiv.org/abs/2309.02081}{{\ttfamily arXiv:2309.02081 [gr-qc]}}.

\bibitem{nakahara2018geometry}
M.~Nakahara, {\em Geometry, topology and physics}.
\newblock CRC press, 2018.

\bibitem{Faddeev:1987ph}
L.~D. Faddeev and L.~A. Takhtajan, {\em {HAMILTONIAN METHODS IN THE THEORY OF
  SOLITONS}}.
\newblock 1987.

\end{thebibliography}\endgroup
\bibliographystyle{utphys}

\end{document}